# The role of ionic liquid breakdown in the electrochemical metallization of $VO_2$: An NMR study of gating mechanisms and $VO_2$ reduction

Michael A. Hope,[1] Kent J. Griffith,[1] Bin Cui,[2] Fang Gao,[2] Siân E. Dutton,[3] Stuart S. P. Parkin,[2] Clare P. Grey[1,*]

[1] Department of Chemistry, University of Cambridge, Lensfield Road, Cambridge CB2 1EW, UK.
[2] Max Planck Institute of Microstructure Physics, Halle (Saale) D06120, Germany.
[3] Cavendish Laboratory, JJ Thomson Avenue, Cambridge CB3 0HE, UK.

## Abstract

Metallization of initially insulating $VO_2$ *via* ionic liquid electrolytes, otherwise known as electrolyte gating, has recently been a topic of much interest for possible applications such as Mott transistors and memory devices. It is clear that the metallization takes place electrochemically and, in particular, there has previously been extensive evidence for the removal of small amounts of oxygen during ionic liquid gating. Hydrogen intercalation has also been proposed, but the source of the hydrogen has remained unclear. In this work, solid-state magic angle spinning NMR spectroscopy ($^1H$, $^2H$, $^{17}O$ and $^{51}V$) is used to investigate the thermal metal-insulator transition in $VO_2$, before progressing to catalytically hydrogenated $VO_2$ and electrochemically metallized $VO_2$. In these experiments electrochemical metallization of bulk $VO_2$ particles is shown to be associated with intercalation of hydrogen, the degree of which can be measured with quantitative $^1H$ NMR spectroscopy. Possible sources of the hydrogen are explored, and by using a selectively deuterated ionic liquid, it is revealed that the hydrogenation is due to deprotonation of the ionic liquid; specifically, for the commonly used dialkylimidazolium based ionic liquids, it is the "carbene" proton that is responsible. Increasing the temperature of the electrochemistry is shown to increase the degree of hydrogenation, forming first a less hydrogenated metallic orthorhombic phase then a more hydrogenated insulating Curie-Weiss paramagnetic orthorhombic phase, both of which were also observed for catalytically hydrogenated $VO_2$. The NMR results are supported by magnetic susceptibility measurements, which corroborate the degree of Pauli and Curie-Weiss paramagnetism. Finally, NMR spectroscopy is used to identify the presence of hydrogen in an electrolyte gated thin film of $VO_2$, suggesting that electrolyte breakdown, proton intercalation and reactions with decomposition products within the electrolyte should not be ignored when interpreting the electronic and structural changes observed in electrochemical gating experiments.

## Introduction

In 1959 it was discovered that upon heating to above 67 °C, vanadium dioxide ($VO_2$) transitions from an insulating to a metallic state with an increase in conductivity of several orders of magnitude;[1] since then, $VO_2$ has been the subject of extensive study to understand the subtle interplay between electronic correlations and a Peierls distortion that underlie this metal-insulator transition (MIT).[2–4] More recently there has been interest in electronically inducing this transition (otherwise known as gating) for possible applications such as Mott transistors[5] and memory devices;[6] this research has focused on thin films of $VO_2$. It was reported by Nakano et al.[7] that non-thermal metallization of $VO_2$ films, induced by application of a gate voltage to an electrolyte at the surface of the film (Figure 1a, left), was a purely capacitive effect, whereby the ionic liquid forms a double layer at the solid–liquid interface and hence induces a large electric field in the sample. Jeong et al.[8] later showed that the metallization was in fact due to the electrochemical reduction of the vanadium and consequent introduction of electrons into the band structure. This reduction must be charge balanced, and Jeong et al. established the simultaneous creation of oxygen vacancies on the basis of $^{18}O$ secondary ion mass spectrometry (SIMS) data, which showed an excess of $^{18}O$ at the surface of devices that had been gated and reverse gated in an $^{18}O_2$ atmosphere; this is the generally accepted mechanism in the literature.[9–14] The same group later showed that oxygen plays a role in ionic liquid gating of several other oxides including $WO_3$, again by $^{18}O$ SIMS.[15] Most recently they directly observed oxygen vacancies, using in-situ transmission electron microscopy, in $SrCoO_{2.5}$ produced by electrolyte gating of $SrCoO_3$, which was accompanied by dramatic structural and magnetic changes.[16] On the other hand, Shibuya and Sawa[17] observed hydrogen intercalation by $^1H$ SIMS after electrolyte gating of $VO_2$, which could also charge balance the reduction; however, the source of hydrogen remained unclear. It is well known that ionic liquids are chemically stable over a limited voltage window and the application of voltages outside this window can lead to breakdown of the organic molecules from which the ionic liquid is composed; furthermore, the possibility of $H_2O$ or other hydrogen-containing impurities in the ionic liquid must also be considered.

Hydrogen can be intercalated into $VO_2$ in the channels parallel to the rutile *c* axis (Figure 1b, bottom); this is clearly a possible mechanism of the metallization of $VO_2$ *via* electrolyte gating because metallization has also been observed after explicit hydrogenation of $VO_2$ by various techniques: (i) electrolytically by splitting of $H_2O$ in a water infiltrated nanoporous glass solid electrolyte[18] or a humid-air nanogap,[19] (ii) galvanically by electrical contact with a sacrificial anode in acidic solution,[20] and (iii) catalytically *via* hydrogen spillover.[21,22] While these studies were all based on nanosized $VO_2$, either thin films or nanowires, hydrogenation of bulk $VO_2$ has also been investigated electrochemically[23] and catalytically,[24] albeit with the studies focusing on the structural rather than the electronic properties. Questions remain, including whether ionic liquid gating is associated with hydrogenation of the $VO_2$? If so, what is the source of hydrogen? And, most importantly, what is the cause of metallization?

Solid-state, magic angle spinning, nuclear magnetic resonance spectroscopy (MAS NMR) is a useful tool to study the metallization of $VO_2$ as it is an element-specific probe of both the crystal and electronic local structures.[25,26] In particular, Knight shifts are a direct measure of the local density of states at the Fermi level for the nucleus in question; they can be used to identify metallic environments and are typically temperature-independent.[27] Paramagnetic shifts due to localized spins, on the other hand, have a strong temperature dependence.[28] These different shift mechanisms can, therefore, be used to determine the local electronic structure.

In this work, multi-nuclear NMR spectroscopy ($^1H$, $^2H$, $^{17}O$ and $^{51}V$) is utilized to study the crystal and electronic structure of electrochemically metallized $VO_2$ samples and explore the mechanism of electrochemical metallization, using the imidazolium-based ionic liquids that are commonly used for electrolyte gating experiments;[8–12] these NMR results are supported by X-ray diffraction (XRD), resistivity and magnetic susceptibility measurements. First, the NMR of pristine micron-sized $VO_2$ particles are investigated above and below the MIT temperature, before the magnetic and electronic properties of catalytically hydrogenated $VO_2$ are examined. The results are compared with electrochemically reduced bulk $VO_2$, and the effect of increasing the temperature at which the electrochemistry is performed is investigated. To explore the source of hydrogen in these experiments, $^2$H NMR measurements were performed on bulk $VO_2$ which was electrochemically reduced with a selectively deuterated ionic liquid; this reveals that deprotonation of the ionic liquid occurs at the voltages used in these experiments, resulting in the observed hydrogenation of $VO_2$. NMR spectroscopy is then used to identify intercalated hydrogen in a thin film sample of $VO_2$ which has been electrolyte-gated with an imidazolium-based ionic liquid, so as to allow comparison between the results obtained with bulk samples and previous thin film electrolyte gating experiments. Finally, the implications for electrolyte gating are discussed.

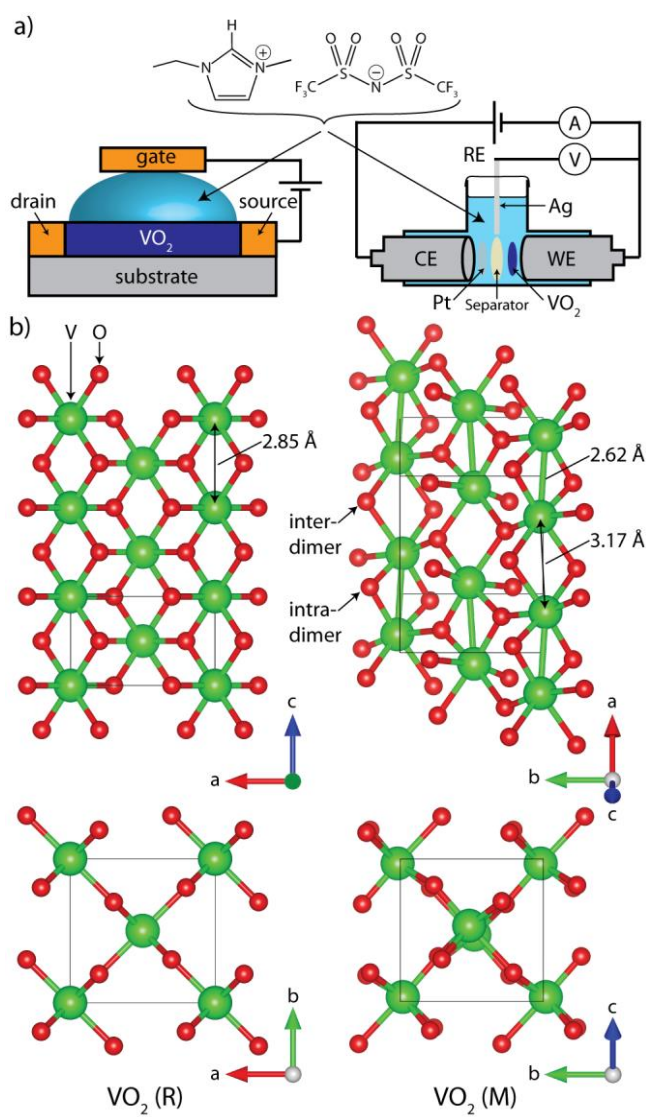

*Figure 1: a) Left: schematic of a thin film electrolyte gating experiment, where a potential is applied across the gate electrode and the resistance between the source and drain is measured. Right: schematic of the three-electrode cell used in this work for electrochemical experiments on bulk $VO_2$ showing the $VO_2$ working electrode (WE), Pt counter electrode (CE) and Ag wire pseudo-reference electrode (RE); shown too is a commonly used ionic liquid, EMIm TFSI. b) Structures of the metallic, rutile, high temperature $VO_2(R)$ phase, and the insulating, monoclinic, low temperature $VO_2(M)$ phase. $VO_2(R)$ has a single oxygen site in the asymmetric unit, while $VO_2(M)$ has two oxygen sites, bridging vanadium atoms either within or between vanadium dimers.*

**Experimental**

*Synthesis:* Bulk $VO_2$ was synthesized by comproportionating an equimolar mixture of $V_2O_3$ and $V_2O_5$ in an evacuated quartz tube at 600 °C for 48 hours to yield ~2 μm particles. The $V_2O_5$ (Sigma-Aldrich, 99.99%) was first dried *in vacuo* at 640 °C for four days and the $V_2O_3$ was synthesized by reducing $V_2O_5$ in 5% $H_2$/Ar (10 mL/min) at 650 °C for 24 hours.[29] $^{17}O$-enriched $VO_2$ was prepared in the same way, but starting from $^{17}O$-enriched $V_2O_5$, which was prepared by oxidizing metallic vanadium powder (Sigma-Aldrich, 99.5%) in 70 at% $^{17}O_2$ gas (Cambridge Isotope Laboratories) at 620 °C for two days.

Catalytically hydrogenated $VO_2$ was prepared by mixing the comproportionated $VO_2$ with Pd nanoparticles (Sciventions, aqueous suspension, 1.5 mg/mL) before removing the water *in vacuo* at 100 °C to give 1 wt% Pd. The Pd/$VO_2$ was then hydrogenated in flowing 25% $H_2$/$N_2$ at 180 °C for 15 hours. A second sample was also prepared by hydrogenation in flowing 5% $H_2$/Ar at 220 °C for 15 hours. Hydrogenated samples were handled in an argon glovebox.

*Characterization:* Powder X-ray diffraction (XRD) patterns were recorded in reflection mode with sample rotation on a PANalytical Empyrean diffractometer emitting Cu Kα (1.540598 Å + 1.544426 Å) radiation. Air-sensitive samples were packed into a Kapton sample holder. Phase identification was achieved by profile matching using PANalytical's X'Pert HighScore Plus 2.2 software and by comparison with the following ICSD entries: 1473 ($V_2O_3$)[30], 74705 ($VO_2$ M),[31] 1504 ($VO_2$ R)[32] and 15798 ($V_2O_5$).[33] Rietveld refinement was performed using the Topas Academic software package.[34] Structures were visualized with the VESTA software package.[35] Thermogravimetric analysis (TGA) was performed under flowing $N_2$ using a Mettler Toledo TGA/SDTA 851 thermobalance with a 100 μL $Al_2O_3$ crucible.

Resistivity measurements were performed on pressed pellets (750 MPa, 30 minutes, under partial vacuum) using the four-point probe technique and a Quantum Design Physical Property Measurement System (PPMS Dynacool). Susceptibility measurements were performed using a Quantum Design Magnetic Property Measurement System (MPMS3) and an applied field of 100 Oe.

Scanning electron microscopy (SEM) was performed using a TESCAN MIRA3 FEG-SEM with an acceleration voltage of 5 kV. The samples were stuck to carbon tape and coated with ~10 nm of Cr. Average particles sizes were determined from the measured images using ImageJ software.[36]

For NMR experiments, samples were packed into $ZrO_2$ rotors. All the NMR spectra were recorded on either a 4.70 T or a 7.05 T Bruker Avance III spectrometer, except one $^2H$ NMR spectrum of $D_xVO_2$ which was recorded on an 11.75 T Bruker Avance III spectrometer. The relatively low magnetic fields used here are advantageous for investigating the NMR of paramagnetic and metallic materials because the paramagnetic and Knight shifts are linear in the applied field, and so constant in chemical shift,[27,28] whereas the sideband separation afforded by magic angle spinning is constant in frequency; greater sideband separation, and hence resolution of signals, can therefore be achieved at lower magnetic fields for the same MAS frequency; furthermore, spinning of metallic samples is easier at lower magnetic fields. Most experiments used a Bruker 1.3 mm HX probe and either 40 kHz or 60 kHz MAS frequency, except the $^2H$ NMR spectra at 4.70 T and 7.05 T which used a Bruker 2.5 mm HX probe and 30 kHz MAS, the $^1H$ NMR spectra of the $VO_2$ thin film which used a Bruker 1.9 mm HX probe and 40 kHz MAS, and the wide temperature range $^1H$ NMR spectra of Pd/$H_xVO_2$ which used a Bruker 4mm HX probe and 14 kHz MAS. All experiments used a Hahn echo pulse sequence unless otherwise stated (π/2-τ-π-τ-acquire). $^1H$ and $^{17}O$ sideband separation experiments were recorded by taking the isotropic slice from a MATPASS experiment,[37] and $^1H$ $T_1$ (spin-lattice) measurements were recorded with an inversion recovery pulse sequence. $^{51}V$ variable offset cumulative spectra (VOCS) were recorded by summing spectra recorded with different carrier frequencies, with retuning of the probe between experiments being performed by an external automatic tuning/matching (eATM) robot.[38] Quantitative $^1H$ NMR spectra were recorded at 4.70 T and 60 kHz MAS, with the sample center-packed between

PTFE tape to ensure excitation of the full sample mass; the integrated intensity was then compared to a calibration with known masses of adamantane, also center-packed. The $T_2$ relaxation constants were sufficiently long that no correction for transverse decay was required. For $^1$H quantification, the catalytically hydrogenated samples were ground with a known mass of KBr to minimize skin depth penetration effects. $^1$H NMR spectra of the VO$_2$ thin film were recorded using a DEPTH background suppression pulse sequence (π/2-τ-π-2τ-π-τ-acquire),[39] and then background-subtracted by first recording the sample then recording the background of an empty rotor with the same experiment and taking the difference. The $T_1$ filtered spectrum was obtained by recording two spectra with recycle delays of 0.05 s and 0.1 s, background-subtracting both, then taking the difference, scaling the spectra so as to minimize the diamagnetic signals which have longer $T_1$ relaxation constants.

Variable temperature NMR experiments were performed by application of heated or cooled nitrogen, with cooling achieved either with a Bruker cooling unit (BCU) or a liquid nitrogen heat exchanger. The sample temperature was determined from an *ex-situ* calibration using the temperature-dependent $^{207}$Pb shift of Pb(NO$_3$)$_2$,[40] except for variable temperature $^1$H spectra of catalytically hydrogenated VO$_2$, which was ground with KBr and the temperature measured *in-situ* from the $^{79}$Br shift and $T_1$ constant.[41] $^1$H NMR spectra were referenced relative to adamantane at 1.81 ppm, $^2$H spectra to D$_2$O at 4.8 ppm, $^{17}$O spectra to CeO$_2$ at 877 ppm and $^{51}$V spectra to NH$_4$VO$_3$ at −571 ppm. Spectra were deconvoluted using the dmfit program.[42]

*Electrochemistry:* Electrochemical experiments were performed with 1-ethyl-3-methylimidazolium bis(trifluoromethylsulfonyl)imide (EMIm TFSI, Sigma Aldrich, ≥97%). The water content was determined with a Metrohm 899 Karl Fischer Coulometer to be 340 ppm as received and 34 ppm after drying *in vacuo* for two days. ½" perfluoroalkoxy (PFA) Swagelok cells were used with a Ag wire pseudo-reference electrode, a platinum mesh counter electrode, a glass fiber separator and stainless-steel plungers (Figure 1a, right).

Composite free-standing films were prepared comprising 80 wt% VO$_2$ particles, 10 wt% PTFE binder and 10 wt% conductive carbon nanoparticles to ensure good electrical contact. VO$_2$ was ground with carbon super P (TIMCAL) before the addition of PTFE (60 wt% dispersion in H$_2$O, Sigma Aldrich). Ethanol was added followed by mixing to a dough-like consistency, rolling and drying at 60 °C to yield films of 75–150 μm thickness. The electrochemical experiments were performed in air using a Bio-Logic potentiostat/galvanostat running the EC-Lab software and experiments at elevated temperatures were performed in an oven. Cells were disassembled in an argon glovebox and the VO$_2$ films washed with dimethyl carbonate (2 x 2.5 ml, 99.5%, anhydrous, Sigma Aldrich) before drying *in vacuo* for 20 minutes. The carbon and PTFE in the composite films make only a small and temperature independent contribution to the magnetic susceptibility.

The potential of the Ag wire pseudo-reference electrode was calibrated relative to the ferrocene–ferrocenium (Fc/Fc$^+$) couple by recording cyclic voltammograms of 10 mM ferrocene in EMIm TFSI at each temperature, with a scan rate of 10 mV s$^{-1}$ (see SI §1). The potential of the reference electrode at 200 °C was extrapolated because ferrocene is not stable at this temperature.[43] The potential *vs.* Fc/Fc$^+$ is related to the potential *vs.* the standard hydrogen electrode (SHE) according to $E - E_{SHE} = E - E_{Fc/Fc^+} + 0.478$ V;[44] the temperature dependence of this conversion is expected to be minimal.[45]

*Thin Films*: Single-crystalline VO$_2$ films of 10×10 mm$^2$ area and around 200 nm thickness were deposited on (001) TiO$_2$ substrates by pulsed laser deposition (248 nm KrF laser) with an oxygen pressure of 0.014 mbar and a growth temperature of 400 °C. The electrolyte gating for the thin film sample was performed potentiostatically according to previously reported procedures,[8] under a vacuum of ~3×10$^{-6}$ mbar at 280 K. The VO$_2$ thin film and a gold counter electrode were covered by a drop of EMIm TFSI and a gate voltage of 3 V was applied between the $VO_2$ thin film and the gold electrode for two hours. After gating, the ionic liquid was removed by ultrasonic cleaning in acetone and ethanol.

**Results and Discussion**

The Thermal Transition of Pure $VO_2$

The high temperature, metallic phase of $VO_2$ adopts the rutile structure ($P4_2/mnm$) with the V $d^1$ electrons delocalized into a conduction band; the transition to the low temperature, insulating, structure is associated with a Peierls distortion to the lower symmetry monoclinic structure ($P2_1/c$), with the V $d^1$ electrons pairing to form V-V dimers (Figure 1b). As has been previously reported, this phase transition results in an extremely large change in shift of the $^{51}V$ NMR signal, from 2065 ppm in the insulating state to −3765 ppm in the metallic state (observed here for our micron-sized $VO_2$ particles, Figure 2a).[25,46] The positive shift is due to a Van-Vleck or orbital Knight shift, which is characteristic of an insulating state with a small bandgap, whereas the negative shift is due to an indirect or core-polarization Knight shift, which is characteristic of a metallic state where the band structure has no appreciable contribution from s orbitals at the Fermi level, as is the case in $VO_2$.[27] The $^{17}O$ NMR spectra (Figure 2b) show a similar effect: above the MIT a negative shift of −505 ppm is observed due to the core-polarization Knight shift of the metallic state, and below the MIT a positive shift is observed due to the Van-Vleck Knight shift of the insulating state. The low temperature spectrum also exhibits a splitting of the $^{17}O$ NMR signal due to the two crystallographically distinct oxygen sites in the lower symmetry monoclinic structure (Figure 1b, right): the peaks at 753 ppm and 814 ppm are tentatively assigned to the inter- and intra- vanadium dimer oxygen environments, respectively, on the basis of preliminary density functional theory (DFT) NMR shielding calculations (see SI, §2). The $^{17}O$ NMR spectrum of $VO_2$ has only previously been reported below the transition,[47] and the two signals were not assigned, but the shifts are in agreement with those found here. Note that the observed $^{17}O$ NMR shifts are not corrected for the second order quadrupolar shift, which from the DFT calculations is expected to contribute around −10 ppm to the observed shift at this field, based on the calculated quadrupolar coupling constants of ~1.6 MHz.

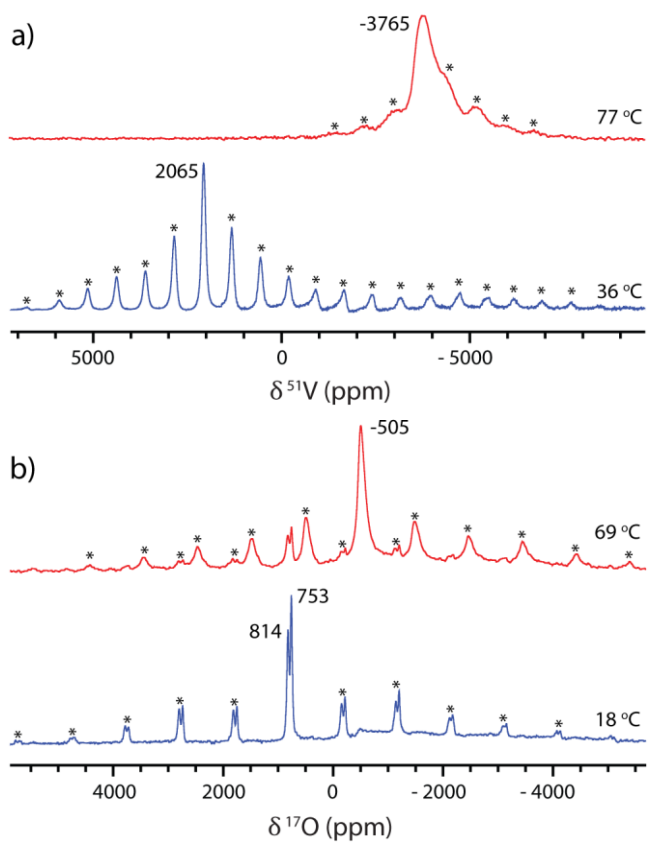

*Figure 2: a) $^{51}$V and b) $^{17}$O NMR spectra of VO$_2$ above and below the MIT, recorded at 7.05 T with a Hahn echo pulse sequence. Spinning sidebands have been marked with an asterisk and the spectra have been scaled to give comparable intensities, rather than being quantitatively comparable. The $^{51}$V NMR spectra were recorded at 60 kHz MAS, summing two spectra recorded with carrier frequencies of 2000 ppm and −4750 ppm. The $^{17}$O NMR spectra were recorded at 40 kHz MAS. Some of the insulating phase remains in the high temperature $^{17}$O NMR spectra, and vice versa, due to spinning induced temperature gradients within the rotor and the finite width of the MIT.*

Catalytic Hydrogenation

In order to explore the crystal and electronic structures of hydrogenated VO$_2$ and the resultant NMR signatures, catalytically hydrogenated VO$_2$ was prepared following the method of Filinchuk et al.[24] *via* catalytic spillover from palladium nanoparticles. Following hydrogenation at 180 °C, a hydrogen content determined from thermogravimetric analysis (TGA) of $x$ = 0.365 was obtained (see SI §3). Rietveld refinement using the powder X-ray diffraction (XRD) pattern showed the presence of a small amount of unreacted VO$_2$ and a mixture of two orthorhombic phases (Pnnm), one with a larger unit cell and orthorhombic distortion than the other (with ratios for the *a* and *b* cell parameters of $b/a$ = 1.112 and 1.036, respectively, see SI §4). Other than the orthorhombic distortion, the orthorhombic phases have the same structure as the high temperature VO$_2$ rutile phase (Figure 1b, left).

This result is in contrast to that of Filinchuk et al. who only found the orthorhombic phase with the smaller unit cell, which they denoted O1, after hydrogenation at 190 °C. However, Chippindale et al.[23] showed that both the size of the unit cell and the orthorhombic distortion scaled with the degree of hydrogenation $x$, suggesting that the second phase identified in our work is a more hydrogenated analogue of the first. Modifying the notation of Filinchuk et al., the less and more hydrogenated orthorhombic phases will be referred to as O1a and O1b respectively. Using the relationship between the orthorhombic distortion and the hydrogen content reported by Chippindale et al.,[23] the hydrogen content of both phases can be predicted, which, combined with the phase fractions determined from Rietveld analysis (see SI §4), results in a total hydrogen content of $x = 0.42(7)$ for this sample (Table 1). This is in reasonable agreement with that determined by TGA.

Four-point resistivity measurements of pressed pellets of VO$_2$ and Pd/H$_x$VO$_2$ (Figure 3a) clearly show the MIT in pristine VO$_2$ at 340 K. This is almost completely suppressed in Pd/H$_x$VO$_2$ (the MIT of the residual unreacted VO$_2$ can, however, still just be seen); furthermore, the resistivity of the Pd/H$_x$VO$_2$ is ~500 times lower than insulating VO$_2$, although the temperature dependence still has semiconducting character rather than being fully metallic (this is most likely due to grain boundary effects). Zero-field cooled susceptibility measurements (Figure 3b) corroborate the resistivity data: the Pd/H$_x$VO$_2$ exhibits an increased temperature-independent susceptibility due to the Pauli paramagnetism of the metallic phase, as well as suppression of the MIT, although there is also an increased Curie paramagnetic component, which is indicative of localized spins.

The $^{51}$V NMR spectrum (Figure 3c) confirms the presence of vanadium atoms in a metallic environment in Pd/H$_x$VO$_2$, with almost complete loss of the insulating VO$_2$ peak at 2065 ppm and the appearance of a resonance at negative shift, as seen for pure VO$_2$ above the MIT; however, the signal in this case is very broad and the spinning sidebands cannot be resolved, which is most likely due to a greater distribution of local vanadium environments in the less uniform H$_x$VO$_2$ sample. The $^1$H MAS NMR spectrum of Pd/H$_x$VO$_2$ contained a series of overlapping signals and thus a MATPASS sideband separation pulse sequence was used so that only the isotropic resonances are seen[37] (Figure 3d); the spectrum shows two signals centered around 110 ppm and 445 ppm, as well as a diamagnetic peak around 0 ppm, which is ascribed to ubiquitous diamagnetic hydrogen-containing impurities.

To aid assignment of the $^1$H spectrum, a second sample of Pd/H$_x$VO$_2$ was synthesized at 220 °C and found from XRD to have a greater phase fraction of O1b (36 wt% *c.f.* 12 wt% for the sample synthesized at 180 °C). The $^1$H NMR spectrum of this sample had a correspondingly greater intensity for the 445 ppm signal, allowing the 115 and 445 ppm regions to be assigned to O1a and O1b respectively (see SI §5).

Variable temperature $^1$H NMR spectra show that the shift of the O1b phase has a Curie-Weiss temperature dependence (see SI §6), indicating that this phase is paramagnetic, the shift originating from localized electrons, with the unpaired electron density in the V t$_{2g}$ orbital partially delocalizing into the H 1s orbital *via* a 90° π delocalization mechanism.[48] These localized electron spins suggest that this phase is insulating, as also recently found for highly catalytically hydrogenated thin films of HVO$_2$ ($x$ = 1);[21] the insulating state in this case is shown to arise from the large degree of hydrogenation which causes the lattice to expand, reducing the overlap between the vanadium d orbitals and hence reducing the valence bandwidth so that there is Mott localization of the electrons, as for the insulating phase of pristine VO$_2$. The O1a resonance, on the other hand, does not show a Curie-Weiss temperature dependence so the major interaction responsible for this signal is likely a Knight shift, indicating that this phase is metallic; this is a positive, direct contact Knight shift because the only valence orbital for hydrogen is the 1s orbital. These assignments are corroborated by measurements of the $^1$H $T_1$ relaxation constants: ~0.03 s for O1a and ~0.002 s for O1b (see SI §7); both phases relax much more quickly than diamagnetic protons (typically ~1–10 s), and the localized paramagnetic O1b signal relaxes an order of magnitude faster than the metallic O1a signal, as expected.

These NMR experiments thus confirm the Curie and Pauli components identified in the magnetic susceptibility measurements. The amount of hydrogenation could also be determined with quantitative $^1$H NMR spectroscopy, which yielded hydrogen contents for the two phases that are in reasonable agreement with the TGA and XRD results (Table 1), further corroborating the assignments.

To summarize, catalytic hydrogenation of VO$_2$ yielded two orthorhombic phases: a less hydrogenated, metallic, Pauli paramagnetic phase denoted O1a; and a more hydrogenated, Curie-Weiss paramagnetic phase denoted O1b; the hydrogen content of both were determined by analysis of the unit cell parameters and by quantitative $^1$H NMR spectroscopy.

Table 1: Comparison of the sample hydrogen content (x in $H_xVO_2$), and its distribution between the two orthorhombic phases, as determined by thermogravimetric analysis (TGA), Rietveld refinement of the XRD pattern and quantitative $^1H$ NMR spectroscopy. The average hydrogen content of the sample is determined from the XRD data by taking the product of the O1a/O1b phase fraction (%) and the phase hydrogen content predicted from the orthorhombic distortion. The error in the last digit is shown in brackets. Discrepancies between the XRD and NMR quantifications are discussed in the SI, see §4.

|       | TGA      | XRD                        | NMR     |
|-------|----------|----------------------------|---------|
| O1a   |          | 86 % × 0.33(5) = 0.28(5)   | 0.19(2) |
| O1b   |          | 12 % × 1.1(1) = 0.14(3)    | 0.17(2) |
| Total | 0.365(3) | 0.42(7)                    | 0.36(3) |

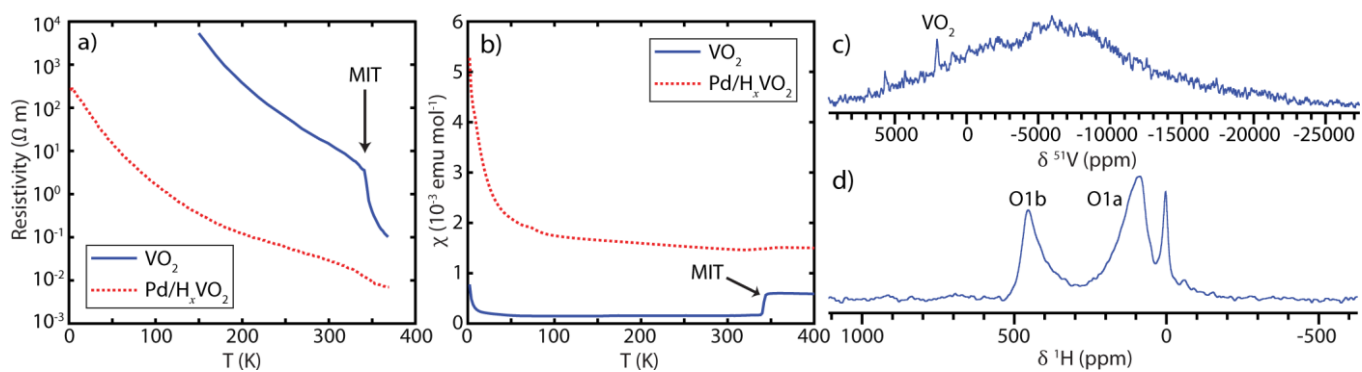

Figure 3: a) Resistivity and b) zero field cooled magnetic susceptibility of $VO_2$ before and after catalytic hydrogenation, and c) $^{51}V$ and d) $^1H$ NMR spectra of the catalytically hydrogenated $VO_2$. The $^{51}V$ NMR spectrum was recorded at 4.70 T, 60 kHz MAS and a sample temperature of ~50 °C, using a Hahn echo pulse sequence and variable offset cumulative spectra (VOCS) acquisition. Spectra were acquired in steps of 5000 ppm between carrier frequencies of 5000 ppm and −20000 ppm and summed to produce the spectrum shown above. The residual signal due to insulating $VO_2$ at 2065 ppm is indicated. The $^1H$ NMR spectrum is the isotropic slice of a 2D MATPASS sideband separation spectrum, which was recorded at 4.70 T, 40 kHz MAS and a sample temperature of ~30 °C.

Electrochemical Hydrogenation – Room Temperature

Having studied the thermal MIT in pure $VO_2$ and the effect of catalytic hydrogenation, electrochemical metallization of $VO_2$ was investigated. The experiments were performed on bulk $VO_2$, using ~15 mg free-standing composite films made with standard battery/supercapacitor electrode preparation techniques. Unlike previous potentiostatic electrolyte gating experiments,[7–14,17] here galvanostatic reduction was used, so that the energetics of different processes could be inferred from the potential, which was measured relative to a silver wire pseudo-reference electrode (Figure 1a, right). The use of the reference electrode avoids electrode polarization effects and allows the potential of the insertion reaction to be measured relative to a known potential, which is particularly important in this case because the reaction that occurs at the counter electrode has not been established definitively. The ionic liquid used was 1-ethyl-3-methylimidazolium bis(trifluoromethylsulfonyl)imide (EMIm TFSI, Figure 1a, top)—a standard electrolyte used in electrolyte gating experiments—and the counter electrode was platinum mesh. A specific current of 6.46 mA g$^{-1}$ was applied for 50 h, which, assuming 100% coulombic efficiency, corresponds to one electron transferred per vanadium atom (Figure 4a).

After performing the electrochemistry, the cell was disassembled under inert atmosphere and the $VO_2$ electrode was characterized *ex-situ*. The presence of protons in a metallic environment is clearly revealed via the observation of a resonance at 110 ppm in the $^1H$ NMR spectrum (Figure 4c); a second peak is observed at approximately 0 ppm, which is ascribed to protons in diamagnetic local environments, from imperfect washing of the electrolyte, electrolyte breakdown products and other hydrogen-containing impurities. Quantification of the $^1$H NMR spectrum, however, yields a hydrogen content of only $x$ = 0.037, despite charge corresponding to one electron per vanadium ion being transferred. Examination of the electrochemistry shows that the electrochemical potential (Figure 4a) initially decreases before reaching a plateau at around −1.6 V *vs.* Fc/Fc$^+$. A second sample was prepared where the electrochemistry was stopped after transferring 0.075 electrons per vanadium, *i.e.* at the beginning of the plateau; this sample had an essentially identical hydrogen content of $x$ = 0.035, which shows that the plateau does not correspond to the hydrogenation reaction, but rather a competing side reaction that prevents further hydrogenation. Electrochemical reduction of VO$_2$ in an organic electrolyte was previously found to compete with hydrogen evolution,[23] *i.e.* the hydrogen evolves as H$_2$ rather than intercalating into the VO$_2$ (the origin of the hydrogen will be discussed later); this is likely to be the case here, given that the voltage falls below the hydrogen evolution voltage in EMIm TFSI (−0.07 V vs. Fc/Fc$^+$), although hydrogen evolution can be negligible until much lower voltages depending on the catalytic properties of the electrode and the source of the hydrogen.[49] Alternative side reactions could also include cation or anion decomposition.[50]

The XRD pattern of the electrochemically reduced VO$_2$ shows that the structure remains monoclinic, but with a lattice expansion consistent with a small degree of hydrogenation (see SI §4). The $^{51}$V NMR spectrum confirms that the monoclinic, insulating, VO$_2$ phase dominates (Figure 4d); the sharp resonance of this phase, at 2065 ppm, and the associated spinning sidebands likely obscure any broad signal due to vanadium in a metallic environment. The susceptibility of this sample (Figure 4b) shows an increase in both the Curie and Pauli paramagnetic susceptibilities relative to VO$_2$, although not as much as for the catalytically hydrogenated sample, and the MIT can still be observed, albeit broadened and suppressed to a lower temperature (332 K on heating *c.f.* 342 K in pristine VO$_2$). The susceptibility also exhibits spin glass-like behavior below $T_f \approx$ 150 K: there is a significant difference between the zero-field cooled (ZFC) and field cooled (FC) susceptibility traces and the ZFC susceptibility increases with temperature below $T_f$ (after the Curie contribution has sufficiently decreased);[51] spin glass-like behavior was further confirmed by hysteresis in the magnetization *vs.* field measurements (see SI §8).

$^{17}$O NMR spectra were recorded on a sample of $^{17}$O-enriched VO$_2$, reduced electrochemically in the same way (Figure 4e). In the spectrum recorded below the MIT temperature, the insulating monoclinic VO$_2$ resonances at 753 and 814 ppm are seen as expected. Then in the spectrum recorded above the MIT, the negatively Knight shifted signal of the metallic phase is again observed, but at the more negative shift of −550 ppm, compared to −505 ppm for pristine metallic VO$_2$ (see SI §9 for a direct comparison of these spectra). This is evidence of the electron (n-type) doping associated with hydrogen intercalation, which increases the density of states at the Fermi level by both raising the Fermi level and increasing the overlap of the V d orbitals; the greater density of states at the Fermi level then increases the magnitude of the Knight shift.

These results suggest localized metallization, as may be expected for low electron doping levels,[4] but not complete metallization; unfortunately, resistivity measurements of these films are not possible due to the conductive carbon and the low density, so the degree of metallization must be inferred. Metallic nanodomains have previously been observed in pure VO$_2$ just below the MIT,[52] and could also explain the behavior observed here for H$_x$VO$_2$ with a low level of hydrogenation: there is a Knight shift for the $^1$H nuclei, indicating that the hydrogen is in a metallic environment, but the whole sample cannot have been metallized because the $^{51}$V NMR spectrum is dominated by vanadium in an insulating environment. Furthermore, metallic nanodomains can also result in cluster glass behavior, with ferromagnetic coupling within domains but weak and disordered coupling between domains, which would explain the spin glass-like effects observed in the magnetic measurements.

To determine whether the galvanostatic experiments performed here would yield the same results as previous two-electrode potentiostatic electrolyte gating experiments, a bulk $VO_2$ sample was electrochemically reduced potentiostatically by applying a voltage of −2.5 V between the Pt counter electrode and the $VO_2$ working electrode (see SI §10). Hydrogen in a metallic environment is again observed by $^1$H NMR, although quantitative NMR yields a lower hydrogen content of $x$ = 0.016; this is ascribed to the smaller applied voltage than in the galvanostatic experiment, where it reaches −3.6 V. Nevertheless, potentiostatic electrochemical reduction of bulk $VO_2$ clearly also results in hydrogenation.

Since the previous electrolyte gating experiments were observed to be reversible,[7,8] a bulk $VO_2$ sample was electrochemically reduced galvanostatically for 24 hours before reversing the current for 24 hours (see SI §11); the $^1$H NMR then shows no $H_xVO_2$ signal, indicating that the electrochemical hydrogenation is also reversible.

The localized/incomplete metallization achieved for bulk $VO_2$ at room temperature is in contrast to the previously reported complete metallization observed for thin film samples. One possible explanation for this difference is the strain present in epitaxial thin films grown on $TiO_2$ (001) substrates; the strain favors the metallic state, as evidenced by the reduction of the thermal MIT temperature to −3 °C.[11] However, for $VO_2$ grown on $Al_2O_3$ ($10\bar{1}0$) substrates there is minimal strain and the MIT is observed at 67 °C as for bulk $VO_2$, yet full metallization is still achieved in electrolyte gating experiments;[8] this suggests that strain is not necessary for metallization. Instead, this could be a kinetic effect: due to the small sample volume in a thin film and the shorter diffusion distances, a greater degree of electrochemical reduction could be achieved before competing side reactions limit the reaction. This is supported by the observation of Passarello et al.[12] that complete suppression of the MIT could not be achieved for 1 µm bars of $VO_2$, whereas it could be achieved for 0.5 µm bars in the same setup; the average $VO_2$ particle size here as determined by SEM is 1.9 µm (see SI §12), which could explain the lack of complete metallization observed. This is corroborated by the results of electrochemical reduction of ∼30 nm $VO_2$ nanoparticles (see SI §13), which show a greater hydrogenation of $x$ = 0.20 from $^1$H NMR.

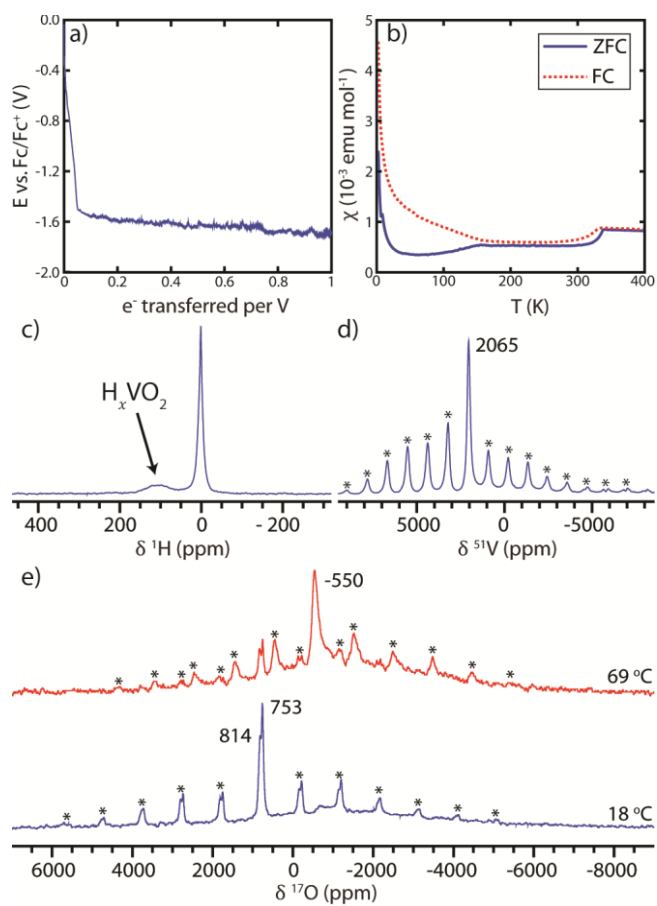

*Figure 4: a) voltage profile; b) zero-field cooled (ZFC) and field cooled (FC) magnetic susceptibilities; and c) $^1$H, d) $^{51}$V and e) $^{17}$O NMR spectra of VO$_2$ after electrochemical reduction at room temperature. The $^1$H NMR spectrum is the isotropic slice of a 2D MATPASS sideband separation spectrum which was recorded at 4.70 T, 40 kHz MAS and a sample temperature of ~30 °C. The $^{51}$V VOCS NMR spectrum was recorded at 4.70 T, 60 kHz MAS and a sample temperature of ~50 °C using a Hahn echo pulse sequence and variable offset cumulative spectra (VOCS) acquisition with carrier frequencies from 5000 ppm to −20000 ppm in steps of 5000 ppm (although only the region of interest is shown here). The $^{17}$O NMR spectra were recorded at 7.05 T, 40 kHz MAS and a sample temperature of ~30 °C. Spinning sidebands are indicated with an asterisk.*

Electrochemical Hydrogenation – Elevated Temperature

The electrochemistry was subsequently performed at elevated temperatures in an attempt to achieve a greater extent of electrochemical hydrogenation. For temperatures up to 150 °C, the degree of hydrogenation increased (Figure 5a), forming the same orthorhombic phases observed for catalytic hydrogenation, first O1a then O1b, as shown by $^1$H NMR spectroscopy (Figure 5c) and Rietveld refinement of the XRD patterns (Figure 5b). The $^{51}$V NMR spectra (Figure 5d) further show a progressive loss of the insulating VO$_2$ resonance at 2065 ppm and the appearance of broad features at negative shifts, which correspond to vanadium ions in a metallic environment. V$^{(0)}$ and V$^{5+}$ impurities can also be seen in the $^{51}$V NMR spectra for the samples electrochemically hydrogenated at 100 °C and 150 °C, which are negligible by XRD but are much more readily observed via $^{51}$V NMR spectroscopy since the signals are noticeably sharper than those of the H$_x$VO$_2$ phases. The susceptibility data corroborate these results; Figure 5e shows the Curie and Pauli paramagnetic components as fitted from the low temperature tail and the high temperature asymptote respectively (see SI §14). As expected, the Pauli paramagnetism increases for samples prepared at up to 150 °C due to the increasing hydrogenation, which is accompanied by the addition of electrons, increasing the density of states at the Fermi level; the Curie paramagnetism also increases for the samples prepared at 100 °C and 150 °C due to the localized paramagnetic O1b phase. The $^{17}$O NMR spectra recorded for $^{17}$O enriched samples also reflect the progressive formation of O1a and then O1b with increasing temperature up to 150 °C, with the paramagnetic O1b phase being identifiable from the Curie-Weiss temperature dependence of the $^{17}$O shift (see SI §15).

The greater electrochemical hydrogenation at higher temperatures could be due to a number of factors. One consideration is that above 67 °C the pristine $VO_2$ is in the metallic rutile phase, which will afford better electrical transport between particles in the electrode as well as presumably reducing the barrier to formation of the orthorhombic phases which are structurally more similar; indeed, no monoclinic phase remains after performing the electrochemistry at 100 °C and above. However, the amount of hydrogenation appears to increase systematically with temperature, rather than there being a step change between the 50 °C and 100 °C samples.

A second explanation is that the differences in the activation energies of the hydrogenation reaction and the limiting side reaction(s) will result in different temperature dependences of the reaction rates; the potential of the plateau in the electrochemistry becomes less negative with increasing temperature (see SI §16), which indicates that the limiting process becomes more facile and hence a change in the kinetics. In particular, the rate of hydrogen diffusion will increase at higher temperatures, reducing the overpotential required to drive the hydrogenation reaction. By 200 °C the degree of hydrogenation is reduced, which is presumably because the competing side reactions have now become faster again relative to the hydrogenation reaction; see SI §17 for a more detailed discussion of this sample.

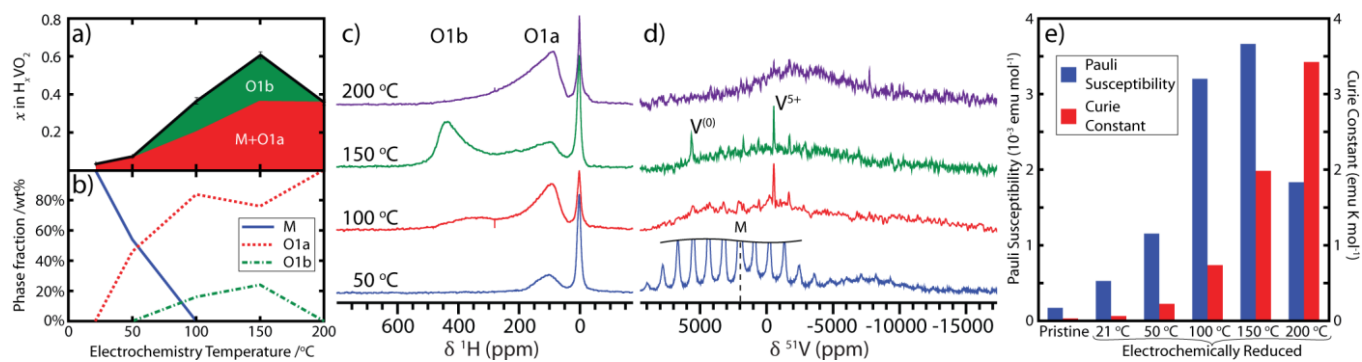

*Figure 5: Characterization of $VO_2$ electrochemically hydrogenated between 50 °C and 200 °C: a) the hydrogen content determined from quantitative $^1$H NMR spectroscopy; b) the phase fractions of the monoclinic (M) and two orthorhombic phases (O1a and O1b) determined from Rietveld analysis of the powder XRD; the c) $^1$H and d) $^{51}$V NMR spectra; and e) the Pauli and Curie paramagnetic components of the magnetic susceptibility. The $^1$H NMR spectra in (c) represent the isotropic slices of the MATPASS spectra which were recorded at 4.70 T, 40 kHz MAS and a sample temperature of ~30 °C; note that these spectra are not quantitative. The $^{51}$V NMR spectra in (d) were recorded at 4.70 T, 60 kHz MAS and a sample temperature of ~50 °C using a Hahn echo pulse sequence and variable offset cumulative spectra (VOCS) acquisition with carrier frequencies from 5000 ppm to −20000 ppm in steps of 5000 ppm. The Pauli component of the susceptibility is taken as the susceptibility measured at 300 K, to avoid any contribution from the MIT, and the Curie constant, C, was found by fitting the low temperature tail to the function $\chi = \frac{C}{T-\Theta} + \chi_0$, where $\Theta$ is the Weiss constant and $\chi_0$ is the temperature-independent susceptibility.*

Electrolyte Gating of Thin Films

To compare the electrochemical metallization experiments on bulk $VO_2$ with the previous studies on thin films, a 200 nm $VO_2$ film was grown on a 0.5 mm $TiO_2$ (001) substrate and electrolyte gated with EMIm TFSI; the film was then crushed and lightly hand ground with a mortar and pestle to allow it to be packed into an NMR sample rotor. As the film cannot be separated from the substrate, there is a 2500-fold dilution of the sample which makes recording the $^1$H NMR spectrum challenging, and the $^{51}$V NMR spectrum essentially impossible using the current substrates. The conventional background-subtracted $^1$H NMR spectrum (Figure 6, top) is dominated by diamagnetic impurities, either from the

TiO$_2$ substrate or the sample surface, obscuring any signal from the gated VO$_2$. However, by applying a $T_1$ filter, the diamagnetic resonances can be largely removed as they relax more slowly (Figure 6, bottom); this leaves signals that relax more quickly, such as metallic H$_x$VO$_2$ environments with $T_1$ ~ 0.03 s. Indeed, in the $T_1$ filtered $^1$H NMR spectrum a resonance can be observed at 115 ppm, where metallic H$_x$VO$_2$ was observed in the electrochemically metallized bulk samples. The same signal was not observed in the $T_1$ filtered $^1$H NMR spectrum of a VO$_2$ thin film before electrolyte gating (see SI §18). Although these experiments are approaching the sensitivity limits of NMR spectroscopy, this suggests that electrolyte gating experiments of thin films also result in hydrogenation of the VO$_2$, due to ionic liquid breakdown.

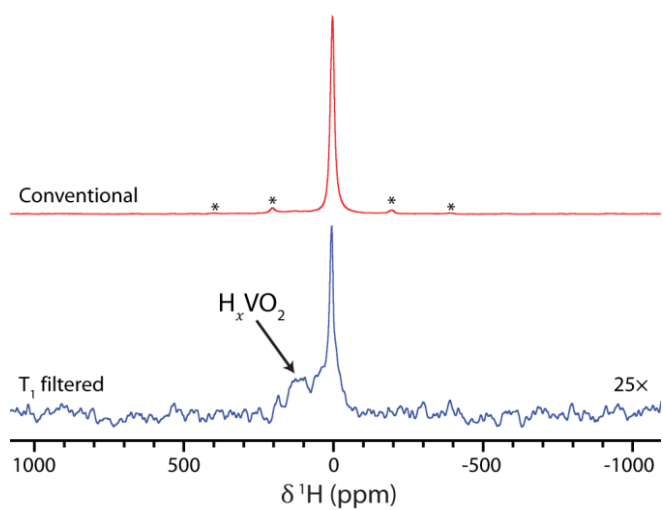

*Figure 6: $^1$H NMR spectra of a 200 nm VO$_2$ thin film on 0.5 mm TiO$_2$ after electrolyte gating, recorded at 4.70 T, 40 kHz MAS and a sample temperature of ~40 °C, with spinning sidebands marked by asterisks. The conventional spectrum was obtained with a recycle delay of 0.05 s using a DEPTH pulse sequence[39] and subtracting the background. The $T_1$ filtered spectrum was recorded by taking the difference between background-subtracted spectra with recycle delays of 0.05 s and 0.1 s, scaling the spectra so as to remove as much as possible the diamagnetic signals. The spectra with recycle delays of 0.05 s were recorded with 2.72 million scans each for the sample and the background, and the spectra with recycle delays of 0.1 s were recorded with 0.68 million scans.*

Electrochemical Hydrogenation – Mechanism

Having established that VO$_2$ can be hydrogenated electrochemically with an ionic liquid electrolyte, the obvious question is: where does the hydrogen come from? One possibility could be electrolysis of H$_2$O which is invariably present in ionic liquids due to their hygroscopic nature,[53] given that hydrogenation of VO$_2$ by water electrolysis has previously been demonstrated.[18,19] However, even after drying the ionic liquid under vacuum for 2 days, electrochemical reduction of VO$_2$ at 100 °C still gave a similar level of hydrogenation, with a greater hydrogen content than can be explained by the water content (1.9 µmol of water in the ionic liquid as determined by Karl Fischer titration, 27 µmol of hydrogen in the electrochemically hydrogenated VO$_2$). Another potential source of hydrogen is the ionic liquid itself, and for 1,3-dialkyl-imidazolium ionic liquids, such as EMIm TFSI, the most acidic proton is the "carbene" proton between the nitrogen atoms of the imidazolium cation, so-called because on deprotonation it forms an N-heterocyclic carbene which is stabilized by the adjacent nitrogen lone pairs[54] (Figure 7a); this deprotonation is driven by the low potential at the VO$_2$ electrode during reduction.[50]

To test this hypothesis, a sample of EMIm TFSI was prepared where the carbene proton had been selectively exchanged for deuterium; this was achieved by stirring EMIm TFSI in excess D$_2$O at 50 °C for 24 hours before drying off the D$_2$O *in vacuo.* A ~90 at% isotopic substitution in the ionic liquid was confirmed by $^1$H and $^2$H NMR spectroscopy (see SI §19). Performing the electrochemical hydrogenation at 100 °C with the selectively deuterated ionic liquid decreased the $^1$H content of the H$_x$VO$_2$ accordingly, as determined by $^1$H NMR spectroscopy. The $^2$H NMR spectrum (Figure 7b) then shows deuterium incorporated in both the O1a and O1b environments, as well as a sharp signal at 0 ppm, due again to diamagnetic decomposition products; the quadrupolar $^2$H nucleus gives rise to a large sideband manifold which can be modelled to find the nuclear quadrupolar coupling constant, see SI §20. This provides compelling evidence that it is the carbene hydrogen of the EMIm TFSI ionic liquid that is intercalated into the $VO_2$ upon electrochemical reduction. The breakdown of the ionic liquid is also evident when removing the ionic liquid after an experiment: for electrochemical reduction at room temperature, the originally clear ionic liquid becomes strongly discolored, and at higher temperatures it becomes dark brown.

Further evidence for this mechanism is seen by using ionic liquids with different imidazolium-based cations, for which the observed potential correlates with the acidity of the protons (see SI §21). For electrolyte gating with non-imidazolium-based ionic liquids, it seems likely that hydrogen intercalation is also involved in the electrochemical metallization of $VO_2$; indeed, for diethylmethyl(2-methoxyethyl)ammonium bis(trifluoromethylsulfonyl)imide (DEME TFSI), another commonly used ionic liquid for electrolyte gating experiments,[7,14,17] hydrogenation is also observed after electrochemical reduction (see SI §22). However, the mechanism of hydrogen abstraction must be different for non-imidazolium-based ionic liquids: this will be the subject of future investigation. To determine the onset voltage of $VO_2$ hydrogenation, bulk $VO_2$ composite films were electrochemically reduced as a function of potential (see SI §23); hydrogenation was first observed at −0.53 V vs. Fc/Fc$^+$, which is much less negative than the reported cathodic stability limit of EMIm TFSI on a glassy carbon electrode, −2.5 V,[55] but this is not surprising given that $VO_2$ both partakes in and catalyses the ionic liquid decomposition.

This work should be contrasted with that of Lu et al. who performed similar electrochemical hydrogenation experiments, using thin films of $SrCoO_{2.5}$ and the ionic liquids EMIm $BF_4$ and DEME TFSI.[56] They added $D_2O$ to the ionic liquids before heating to 100 °C and then performed the electrochemistry; they subsequently observed $^2$H ions from SIMS in the gated material and therefore concluded that the hydrogen arose from $H_2O$ ($D_2O$) in the ionic liquid. Although this is a different material, it is possible that the heating caused exchange of the labile proton on the ionic liquid for deuterium, and hence that the hydrogenation is also due to decomposition of the ionic liquid in this case. However, further investigation would be required to unambiguously determine the source of hydrogen in this different system, in particular because the potential at which the oxide is reduced may be important in determining the mechanism of hydrogenation.

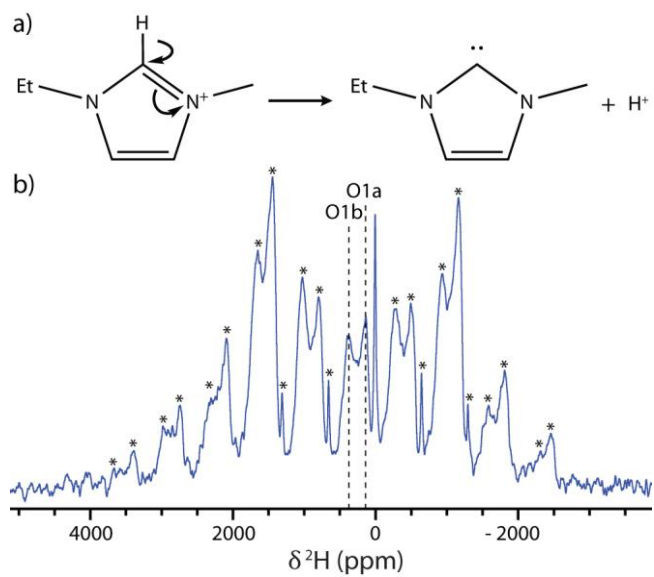

*Figure 7: a) Deprotonation of the "carbene hydrogen" in the EMIm cation. b) The $^2$H NMR spectrum of VO$_2$ after electrochemical metallization with EMIm TFSI which has had the carbene hydrogen exchanged for deuterium, showing signals from both O1a and O1b D$_x$VO$_2$. The spectrum was recorded at 7.05 T, 30 kHz MAS and a sample temperature of ~40 °C using a Hahn echo pulse sequence. Spinning sidebands are shown with asterisks.*

The experiments on the electrochemical reduction of VO$_2$ presented here do not rule out the presence of oxygen vacancies, but they suggest that hydrogenation could be a sufficient explanation for the metallization, particularly for certain classes of ionic liquids and if large overpotentials (gating voltages) are used. The electrochemical reaction of the VO$_2$ in this case is

$$\text{VO}_2 + x\text{H}^+ + xe^- \rightarrow \text{H}_x\text{VO}_2 \quad [1]$$

However, even in systems where oxygen vacancies dominate, such as WO$_3$ and SrCoO$_{2.5}$,[15,16,56] we propose that hydrogen still plays an important role in the electrochemistry; this hydrogen would most likely come from decomposition of the ionic liquid, although as discussed above, the balance between water and electrolyte decomposition may depend on both the system and the applied potential. The charge balancing of metal reduction in the metal oxide MO$_n$ by loss of oxygen could be written as

$$\text{MO}_n \rightarrow \text{MO}_{n-x} + \frac{x}{2}\text{O}_2 \quad [2]$$

However, this above formulation of the process is too simplistic because it is a purely chemical reaction and O$_2$ cannot be generated at the negative electrode (where the reduction occurs). Instead the proposed reaction must occur by two electrochemical half-reactions, at the negative (metal oxide) and positive (gate/counter) electrodes respectively:

$$\text{MO}_n + 2xe^- \rightarrow \text{MO}_{n-x} + x\text{O}^{2-} \quad \text{(Negative electrode)} \quad [3]$$

$$x\text{O}^{2-} \rightarrow \frac{x}{2}\text{O}_2 + 2xe^- \quad \text{(Positive electrode)} \quad [4]$$

Transport of O$^{2-}$ between the electrodes is formally required, which we suggest could occur as H$_2$O, formed from the protons liberated *via* electrolyte decomposition, rather than as a free O$^{2-}$ ion. *I.e.*

$$\text{MO}_n + 2x\text{H}^+ + 2xe^- \rightarrow \text{MO}_{n-x} + x\text{H}_2\text{O} \quad \text{(Negative electrode)} \quad [5]$$

$$x\text{H}_2\text{O} \rightarrow \frac{x}{2}\text{O}_2 + 2x\text{H}^+ + 2xe^- \quad \text{(Positive electrode)} \quad [6]$$

This reaction has direct analogies with the onset of so-called conversion reactions in lithium-ion battery electrodes where $Li_2O$ is generated along with the reduction of the metal ions, eventually to the metal (e.g. $CoO + 2Li^+ + 2e^- \rightarrow Co + Li_2O$).[57,58] Note that these conversion reactions can also commence with lithiation (intercalation) before conversion, which is again analogous to the proton intercalation observed here in $VO_2$.

Conclusions

Electrochemical metallization of micron-sized $VO_2$ particles with imidazolium ionic liquids has been shown to be associated with intercalation of protons and concomitant reduction of the $V^{4+}$ ions, $^1H$ NMR spectra with a positive Knight shift due to the metallization providing a clear signature of this event. There is also evidence for the same hydrogenation in thin films of $VO_2$, after light grinding of the film into small pieces so as to pack into the sample container. In the case of 1,3-dialkyl-imidazolium-based ionic liquids, which are common for previously reported electrolyte gating experiments, the hydrogenation is due to deprotonation of the ionic liquid, specifically the "carbene" hydrogen of the imidazolium cation; this has been shown by selectively substituting this hydrogen for deuterium.

Electrochemical reduction of bulk $VO_2$ at room temperature does not afford complete metallization, but rather localized metallization in the vicinity of the intercalated H, which is in contrast to thin film electrolyte gating experiments; greater hydrogenation could, however, be achieved for nanoparticulate $VO_2$. Increasing the temperature of the electrochemistry also yields greater hydrogenation, forming first a metallic orthorhombic phase and then a second localized paramagnetic orthorhombic phase with a greater degree of hydrogenation; a schematic phase diagram for $H_xVO_2$ is shown in Figure 8. A mixture of the same orthorhombic phases was also observed for catalytically hydrogenated $VO_2$, for which the resistivity was shown to decrease by a factor of 500 compared to pristine $VO_2$. The degree of hydrogenation can be measured by quantitative $^1H$ NMR spectroscopy, and the Pauli and Curie paramagnetic components of the two orthorhombic phases can be tracked *via* $^1H$, $^{17}O$ and $^{51}V$ NMR spectroscopy; in particular, the $^{17}O$ Knight shift in the metallic phase is a sensitive probe of the density of states at the Fermi level and hence the degree of electron doping, and variable temperature experiments for both $^1H$ and $^{17}O$ can be used to assign the Curie paramagnetic phases.

These results should be taken into consideration when developing a device based on electrolyte gating of $VO_2$ thin films: the carbene species formed on deprotonation of the ionic liquid is very reactive and will cause decomposition of the electrolyte and degradation of the performance of the device. However, in systems where hydrogenation is the cause of the metallization, an alternative electrochemical system can be formulated to intentionally and reversibly intercalate hydrogen, which could allow the practical realization of electrolyte gating in devices. Finally, we suggest that the protons produced by electrolyte degradation may be involved in oxygen extraction mechanisms in this and other electrochemically-gated systems, a proposal that is currently under investigation.

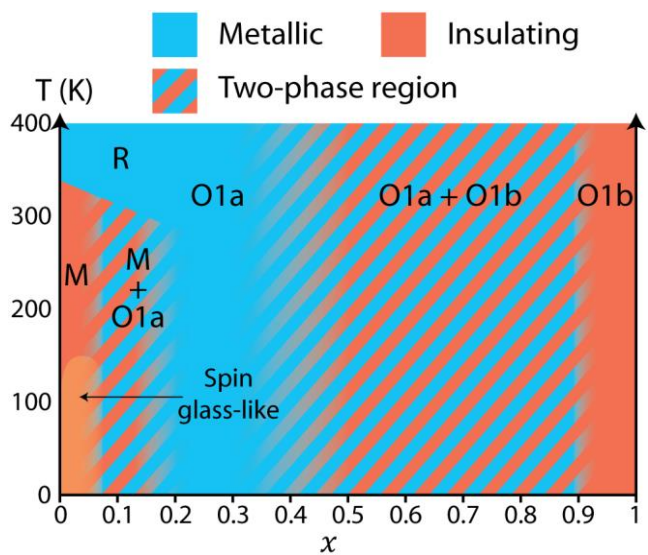

*Figure 8: Schematic phase diagram of $H_xVO_2$, showing the monoclinic (M), rutile (R) and two orthorhombic phases (O1a and O1b), with two phase regions cross hatched. Boundaries were estimated from the compositions of the $H_xVO_2$ samples studied in this work, rather than being rigorously mapped. The M and O1b phases are insulating while the R and O1a phases are metallic. The single-phase rutile (R) region for $x \approx 0.1$ above the MIT temperature was determined by variable temperature XRD, see SI §24.*

**Acknowledgements**

M.A.H. would like to thank the Oppenheimer Foundation for funding. K.J.G. gratefully acknowledges support from The Winston Churchill Foundation of the United States, the Herchel Smith Scholarship and the EPSRC (EP/M009521/1). This paper is part of a project that has received funding from the European Union's Horizon 2020 research and innovation program under grant agreement No. 737109. B.C. thanks the Alexander von Humboldt Foundation for their support. Finally, we would like to thank all the members of the Grey Group who provided help, advice and discussion for this work.

*Supporting information.*
Additional experiments as referenced in the text are supplied as Supporting Information.

Experimental NMR, magnetic susceptibility, and XRD data have been made available at
[doi.org/10.17863/CAM.32043](doi.org/10.17863/CAM.32043).

References

(1)   Morin, F. J. Oxides Which Show a Metal-to-Insulator Transition at the Neel Temperature. *Phys. Rev. Lett.* **1959**, *3* (1), 34–36.

(2)   Goodenough, J. B. The Two Components of the Crystallographic Transition in VO2. *J. Solid State Chem.* **1971**, *3* (4), 490–500.

(3)   Rice, T. M.; Launois, H.; Pouget, J. P. Comment on "VO2: Peierls or Mott-Hubbard? A View from Band Theory." *Phys. Rev. Lett.* **1994**, *73* (22), 3042–3042.

(4)   Khomskii, D. I. *Transition Metal Compounds*; Cambridge University Press: Cambridge, 2014.

(5)   Chudnovskiy, F.; Luryi, S.; Spivak, B. Switching Device Based on First-Order Metal- Insulator Transition Induced by External Electric Field. *Futur. Trends Microelectron. Nano Millenn.* **2002**, 148–155.

(6)   Driscoll, T.; Kim, H. T.; Chae, B. G.; Di Ventra, M.; Basov, D. N. Phase-Transition Driven Memristive System. *Appl. Phys. Lett.* **2009**, *95* (4), 93–96.

(7) Nakano, M.; Shibuya, K.; Okuyama, D.; Hatano, T.; Ono, S.; Kawasaki, M.; Iwasa, Y.; Tokura, Y. Collective Bulk Carrier Delocalization Driven by Electrostatic Surface Charge Accumulation. *Nature* **2012**, *487* (7408), 459–462.

(8) Jeong, J.; Aetukuri, N.; Graf, T.; Schladt, T. D.; Samant, M. G.; Parkin, S. S. P. Suppression of Metal-Insulator Transition in VO2 by Electric Field-Induced Oxygen Vacancy Formation. *Science* **2013**, *339* (6126), 1402–1405.

(9) Chen, S.; Wang, X. J.; Fan, L.; Liao, G.; Chen, Y.; Chu, W.; Song, L.; Jiang, J.; Zou, C. The Dynamic Phase Transition Modulation of Ion-Liquid Gating VO2 Thin Film: Formation, Diffusion, and Recovery of Oxygen Vacancies. *Adv. Funct. Mater.* **2016**, *26* (20), 3532–3541.

(10) Gupta, S. N.; Pal, A.; Muthu, D. V. S.; Anil Kumar, P. S.; Sood, A. K. Metallic Monoclinic Phase in VO2 Induced by Electrochemical Gating: In Situ Raman Study. *EPL* **2016**, *115* (1), 17001.

(11) Jeong, J.; Aetukuri, N. B.; Passarello, D.; Conradson, S. D.; Samant, M. G.; Parkin, S. S. P. Giant Reversible, Facet-Dependent, Structural Changes in a Correlated-Electron Insulator Induced by Ionic Liquid Gating. *Proc. Natl. Acad. Sci.* **2015**, *112* (4), 1013–1018.

(12) Passarello, D.; Altendorf, S. G.; Jeong, J.; Rettner, C.; Arellano, N.; Topuria, T.; Samant, M. G.; Parkin, S. S. P. Evidence for Ionic Liquid Gate-Induced Metallization of Vanadium Dioxide Bars over Micron Length Scales. *Nano Lett.* **2017**, *17* (5), 2796–2801.

(13) Dahlman, C. J.; LeBlanc, G.; Bergerud, A.; Staller, C.; Adair, J.; Milliron, D. J. Electrochemically Induced Transformations of Vanadium Dioxide Nanocrystals. *Nano Lett.* **2016**, *16* (10), 6021–6027.

(14) Singh, S.; Abtew, T. A.; Horrocks, G.; Kilcoyne, C.; Marley, P. M.; Stabile, A. A.; Banerjee, S.; Zhang, P.; Sambandamurthy, G. Selective Electrochemical Reactivity of Rutile VO2 towards the Suppression of Metal-Insulator Transition. *Phys. Rev. B* **2016**, *93* (12), 1–8.

(15) Altendorf, S. G.; Jeong, J.; Passarello, D.; Aetukuri, N. B.; Samant, M. G.; Parkin, S. S. P. Facet-Independent Electric-Field-Induced Volume Metallization of Tungsten Trioxide Films. *Adv. Mater.* **2016**, *28* (26), 5284–5292.

(16) Cui, B.; Werner, P.; Ma, T.; Zhong, X.; Wang, Z.; Taylor, J. M.; Zhuang, Y.; Parkin, S. S. P. Direct Imaging of Structural Changes Induced by Ionic Liquid Gating Leading to Engineered Three-Dimensional Meso-Structures. *Nat. Commun.* **2018**, *9* (1), 3055.

(17) Shibuya, K.; Sawa, A. Modulation of Metal-Insulator Transition in VO2 by Electrolyte Gating-Induced Protonation. *Adv. Electron. Mater.* **2016**, *2* (2), 1500131.

(18) Katase, T.; Endo, K.; Tohei, T.; Ikuhara, Y.; Ohta, H. Room-Temperature-Protonation-Driven On-Demand Metal-Insulator Conversion of a Transition Metal Oxide. *Adv. Electron. Mater.* **2015**, *1* (7), 1500063.

(19) Sasaki, T.; Ueda, H.; Kanki, T.; Tanaka, H. Electrochemical Gating-Induced Reversible and Drastic Resistance Switching in VO2 Nanowires. *Sci. Rep.* **2015**, *5* (1), 17080.

(20) Chen, Y.; Wang, Z.; Chen, S.; Ren, H.; Wang, L.; Zhang, G.; Lu, Y.; Jiang, J.; Zou, C.; Luo, Y. Non-Catalytic Hydrogenation of VO2 in Acid Solution. *Nat. Commun.* **2018**, *9* (1), 818.

(21) Yoon, H.; Choi, M.; Lim, T.-W.; Kwon, H.; Ihm, K.; Kim, J. K.; Choi, S.-Y.; Son, J. Reversible Phase Modulation and Hydrogen Storage in Multivalent VO2 Epitaxial Thin Films. *Nat. Mater.* **2016**, *15* (10), 1113–1119.

(22) Wei, J.; Ji, H.; Guo, W.; Nevidomskyy, A. H.; Natelson, D. Hydrogen Stabilization of Metallic Vanadium Dioxide in Single-Crystal Nanobeams. *Nat. Nanotechnol.* **2012**, *7* (6), 357–362.

(23) Chippindale, A. M.; Dickens, P. G.; Powell, A. V. Synthesis, Characterization, and Inelastic Neutron Scattering Study of Hydrogen Insertion Compounds of VO2(Rutile). *J. Solid State Chem.* **1991**, *93* (2), 526–533.

(24) Filinchuk, Y.; Tumanov, N. A.; Ban, V.; Ji, H.; Wei, J.; Swift, M. W.; Nevidomskyy, A. H.; Natelson, D. In Situ Diffraction Study of Catalytic Hydrogenation of VO2 : Stable Phases and Origins of Metallicity. *J. Am. Chem. Soc.* **2014**, *136* (22), 8100–8109.

(25) Gro Nielsen, U.; Skibsted, J.; Jakobsen, H. J. β-VO2—a V(IV) or a Mixed-Valence V(III)–V(V) Oxide—studied by 51V MAS NMR Spectroscopy. *Chem. Phys. Lett.* **2002**, *356* (1–2), 73–78.

(26) Lynch, G. .; Segel, S. .; Sayer, M. Nuclear Magnetic Resonance Study of Polycrystalline VO2. *J. Magn. Reson.* **1974**, *15* (1), 8–18.

(27) Bennett, L. H.; Watson, R. E.; Carter, G. C. Relevance of Knight Shift Measurements to the Electronic Density of States. *J. Res. Natl. Bur. Stand. Sect. A Phys. Chem.* **1970**, *74A* (4), 569.

(28) Pell, A. J.; Pintacuda, G.; Grey, C. P. Paramagnetic NMR in Solution and the Solid State. *Prog. Nucl. Magn. Reson. Spectrosc. (in press)*.

(29) *Gmelins Handbuch Der Anorganischen Chemie, Vanadium, Vol: B1*, 8th ed.; Chemie gmbh weinheim/bergstr, 1967.

(30) Rice, C. E.; Robinson, W. R. Structural Changes in the Solid Solution (Ti1−xVx)2O3 as x Varies from Zero to One. *J. Solid State Chem.* **1977**, *21* (2), 145–154.

(31) Rogers, K. D. An X-Ray Diffraction Study of Semiconductor and Metallic Vanadium Dioxide. *Powder Diffr.* **1993**, *8* (4), 240–244.

(32) Ghedira, M.; Vincent, H.; Marezio, M.; Launay, J. C. Structural Aspects of the Metal-Insulator Transitions in V0.985Al0.015O2. *J. Solid State Chem.* **1977**, *22* (4), 423–438.

(33) Bachmann, H.G.; Ahmed, F.R.; Barnes, W. H. The Crystal Structure of Vanadium Pentoxide. *Zeitschrift fuer Krist. Krist. Krist. Krist.* **1961**, *115*, 110–131.

(34) Coelho, A. A. Indexing of Powder Diffraction Patterns by Iterative Use of Singular Value Decomposition. *J. Appl. Crystallogr.* **2003**, *36* (1), 86–95.

(35) Momma, K.; Izumi, F. VESTA 3 for Three-Dimensional Visualization of Crystal, Volumetric and Morphology Data. *J. Appl. Crystallogr.* **2011**, *44* (6), 1272–1276.

(36) Schneider, C. A.; Rasband, W. S.; Eliceiri, K. W. NIH Image to ImageJ: 25 Years of Image Analysis. *Nat. Methods* **2012**, *9* (7), 671.

(37) Hung, I.; Zhou, L.; Pourpoint, F.; Grey, C. P.; Gan, Z. Isotropic High Field NMR Spectra of Li-Ion Battery Materials with Anisotropy >1 MHz. *J. Am. Chem. Soc.* **2012**, *134* (4), 1898–1901.

(38) Pecher, O.; Halat, D. M.; Lee, J.; Liu, Z.; Griffith, K. J.; Braun, M.; Grey, C. P. Enhanced Efficiency of Solid-State NMR Investigations of Energy Materials Using an External Automatic Tuning/Matching (EATM) Robot. *J. Magn. Reson.* **2017**, *275*, 127–136.

(39) Robin Bendall, M.; Gordon, R. E. Depth and Refocusing Pulses Designed for Multipulse NMR with Surface Coils. *J. Magn. Reson.* **1983**, *53* (3), 365–385.

(40) Bielecki, A.; Burum, D. P. Temperature Dependence of 207 Pb MAS Spectra of Solid Lead Nitrate. An Accurate, Sensitive Thermometer for Variable-Temperature MAS. *J. Magn. Reson. Ser. A* **1995**, *116*, 215–220.

(41) Thurber, K. R.; Tycko, R. Measurement of Sample Temperatures under Magic-Angle Spinning from the Chemical Shift and Spin-Lattice Relaxation Rate of 79Br in KBr Powder. *J. Magn.*

*Reson.* **2009**, *196* (1), 84–87.

(42) Massiot, D.; Fayon, F.; Capron, M.; King, I.; Le Calvé, S.; Alonso, B.; Durand, J.-O.; Bujoli, B.; Gan, Z.; Hoatson, G. Modelling One- and Two-Dimensional Solid-State NMR Spectra. *Magn. Reson. Chem.* **2002**, *40* (1), 70–76.

(43) De Souza, A. C.; Pires, A. T. N.; Soldi, V. Thermal Stability of Ferrocene Derivatives and Ferrocene-Containing Polyamides. *J. Therm. Anal. Calorim.* **2002**, *70* (2), 405–414.

(44) Bizzarri, C.; Conte, V.; Floris, B.; Galloni, P. Solvent Effects of Ionic Liquids: Investigation of Ferrocenes as Electrochemical Probes. *J. Phys. Org. Chem.* **2011**, *24* (4), 327–334.

(45) Matsumiya, M.; Terazono, M.; Tokuraku, K. Temperature Dependence of Kinetics and Diffusion Coefficients for Ferrocene/Ferricenium in Ammonium-Imide Ionic Liquids. *Electrochim. Acta* **2006**, *51* (7), 1178–1183.

(46) Umeda, J. J.; Kusumoto, H.; Narita, K.; Yamada, E. Nuclear Magnetic Resonance in Polycrystalline VO2. *J. Chem. Phys.* **1965**, *42* (1965), 1458.

(47) Bastow, T. J.; Stuart, S. N. 17O NMR in Simple Oxides. *Chem. Phys.* **1990**, *143* (3), 459–467.

(48) Carlier, D.; Ménétrier, M.; Grey, C. P.; Delmas, C.; Ceder, G. Understanding the NMR Shifts in Paramagnetic Transition Metal Oxides Using Density Functional Theory Calculations. *Phys. Rev. B* **2003**, *67* (17), 174103.

(49) Meng, Y.; Aldous, L.; Belding, S. R.; Compton, R. G. The Hydrogen Evolution Reaction in a Room Temperature Ionic Liquid: Mechanism and Electrocatalyst Trends. *Phys. Chem. Chem. Phys.* **2012**, *14* (15), 5222.

(50) DeVos, N.; Maton, C.; Stevens, C. V. Electrochemical Stability of Ionic Liquids: General Influences and Degradation Mechanisms. *ChemElectroChem* **2014**, *1* (8), 1258–1270.

(51) Mydosh, J. A. *Spin Glasses: An Experimental Introduction*; Taylor & Francis: London, 1993.

(52) Qazilbash, M. M.; Brehm, M.; Chae, B.-G.; Ho, P.-C.; Andreev, G. O.; Kim, B.-J.; Yun, S. J.; Balatsky, A. V; Maple, M. B.; Keilmann, F.; Kim, H.-T.; Basov, D. N. Mott Transition in VO2 Revealed by Infrared Spectroscopy and Nano-Imaging. *Science* **2007**, *318* (5857), 1750–1753.

(53) Krannich, M.; Heym, F.; Jess, A. Characterization of Six Hygroscopic Ionic Liquids with Regard to Their Suitability for Gas Dehydration: Density, Viscosity, Thermal and Oxidative Stability, Vapor Pressure, Diffusion Coefficient, and Activity Coefficient of Water. *J. Chem. Eng. Data* **2016**, *61* (3), 1162–1176.

(54) Hollóczki, O.; Gerhard, D.; Massone, K.; Szarvas, L.; Németh, B.; Veszprémi, T.; Nyulászi, L. Carbenes in Ionic Liquids. *New J. Chem.* **2010**, *34* (12), 3004.

(55) Mousavi, M. P. S.; Dittmer, A. J.; Wilson, B. E.; Hu, J.; Stein, A.; Bühlmann, P. Unbiased Quantification of the Electrochemical Stability Limits of Electrolytes and Ionic Liquids. *J. Electrochem. Soc.* **2015**, *162* (12), A2250–A2258.

(56) Lu, N.; Zhang, P.; Zhang, Q.; Qiao, R.; He, Q.; Li, H.-B.; Wang, Y.; Guo, J.; Zhang, D.; Duan, Z.; Li, Z.; Wang, M.; Yang, S.; Yan, M.; Arenholz, E.; Zhou, S.; Yang, W.; Gu, L.; Nan, C.-W.; Wu, J.; Tokura, Y.; Yu, P. Electric-Field Control of Tri-State Phase Transformation with a Selective Dual-Ion Switch. *Nature* **2017**, *546* (7656), 124–128.

(57) Poizot, P.; Laruelle, S.; Grugeon, S.; Dupont, L.; Tarascon, J.-M. Nano-Sized Transition-Metal Oxides as Negative-Electrode Materials for Lithium-Ion Batteries. *Nature* **2000**, *407* (6803), 496–499.

(58) Yamakawa, N.; Jiang, M.; Grey, C. P. Investigation of the Conversion Reaction Mechanisms for

Binary Copper(II) Compounds by Solid-State NMR Spectroscopy and X-Ray Diffraction. *Chem. Mater.* **2009**, *21* (14), 3162–3176.

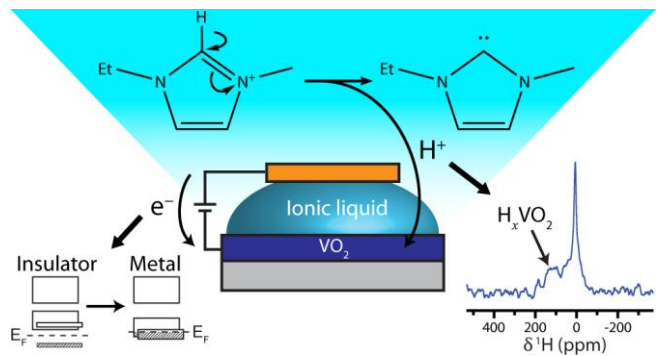

Supporting Information for

# The role of ionic liquid breakdown in the electrochemical metallization of VO$_2$: An NMR study of gating mechanisms and VO$_2$ reduction

Michael A. Hope,[1] Kent J. Griffith,[1] Bin Cui,[2] Fang Gao,[2] Siân E. Dutton,[3] Stuart S. P. Parkin,[2] Clare P. Grey[1,*]

[1] Department of Chemistry, University of Cambridge, Lensfield Road, Cambridge CB2 1EW, UK.
[2] Max Planck Institute of Microstructure Physics, Halle (Saale) D06120, Germany.
[3] Cavendish Laboratory, JJ Thomson Avenue, Cambridge CB3 0HE, UK.

# Contents

# 1 Calibration of the Ag Wire Pseudo-Reference Electrode

To calibrate the Ag wire pseudo-reference electrode, cyclic voltammograms (CVs) were recorded of 10 mM ferrocene (Fc) dissolved in the EMIm TFSI ionic liquid electrolyte, at room temperature and in steps of 25 °C between 50 °C and 175 °C, inclusive (Figure 1a). The potential of the pseudo-reference electrode relative to the ferrocene/ferrocenium (Fc/Fc$^+$) redox couple can then be found at each temperature from the (negative of the) average of the anodic and cathodic peak potentials in the CVs (Figure 1b). The potential of the Ag wire pseudo-reference electrode increases with increasing temperature due to greater solubility of Ag$^+$ in the ionic liquid. The ferrocene/ferrocenium couple cannot be measured at 200 °C as ferrocene decomposes at this temperature,[1] so the potential of the reference electrode was extrapolated to this temperature.

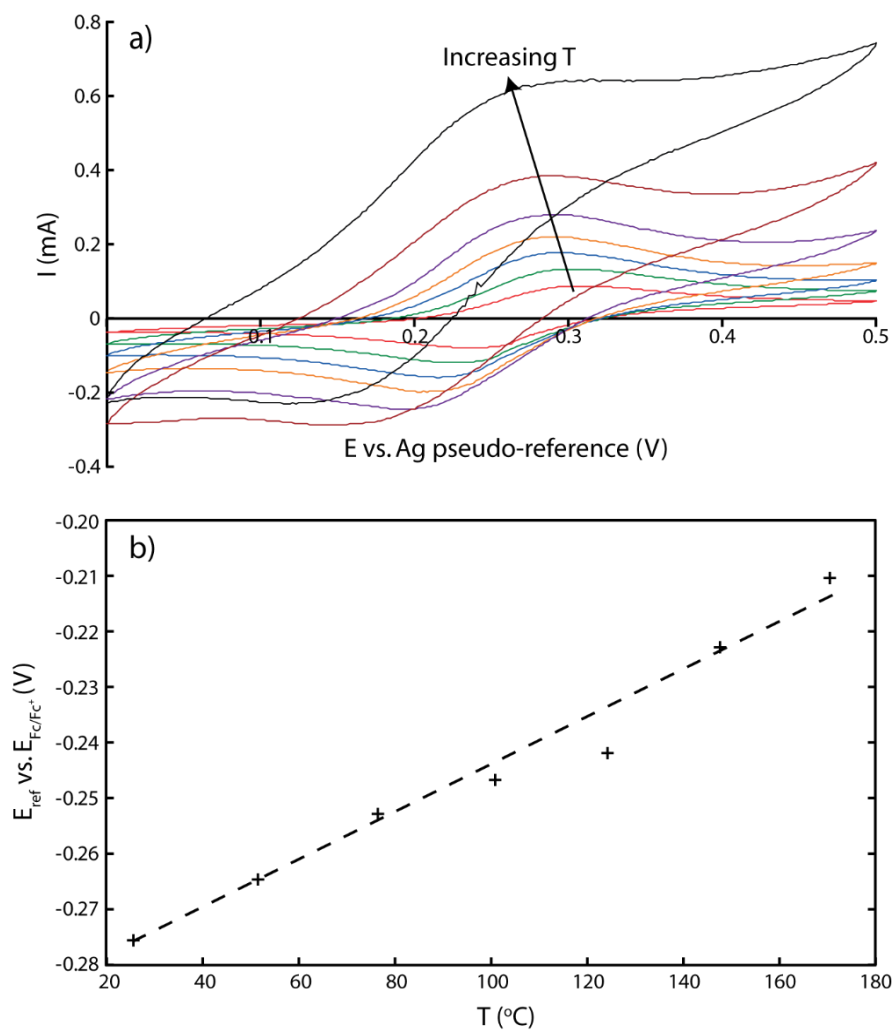

*Figure 1: a) Cyclic voltammograms of 10 mM ferrocene in EMIm TFSI measured relative to a Ag wire pseudo-reference electrode, recorded at room temperature and in steps of 25 °C between 50 °C and 175 °C with a sweep rate of 10 mV/s. b) The potential of the Ag pseudo-reference electrode at each temperature, measured relative to the ferrocene/ferrocenium couple, as determined from the average of the peak potentials in the cyclic voltammograms.*

# 2 DFT Calculations

The correlated nature of the electrons in VO$_2$ makes the use of DFT to calculate the electronic structure challenging, and a number of approaches have previously been applied to accurately study VO$_2$, including Hubbard U corrections,[2] hybrid functionals[3] and dynamical mean field theory.[4] However, calculations of magnetic resonance properties using the Gauge Including Projector Augmented Waves (GIPAW) method are either too computationally expensive or not currently implemented with these methods; instead, for the purposes of spectral assignment and estimation of quadrupolar parameters, a generalised gradient approximation (GGA) approach was used here.

DFT calculations of NMR parameters were performed using the CASTEP plane wave density functional theory (DFT) code[5–8] and the PBE exchange-correlation functional,[9] with a plane wave cut-off energy of 700 eV and a 6 × 6 × 6 Monkhorst–Pack *k*-point mesh. The relationship between the calculated chemical shielding and the experimental chemical shift was determined by calculating the isotropic value of the $^{17}$O NMR shielding tensor for selected diamagnetic binary first-row transition metal oxides—$Sc_2O_3$, rutile $TiO_2$ and $V_2O_5$ (the last of which has three different oxygen sites)—and plotting this against the experimental isotropic chemical shift[10] (Figure 2). The atomic positions and unit cell parameters were relaxed from the experimental structures, which were taken from ICSD entries 26841,[11] 62677[12] and 15798,[13] respectively.

The calculation of the NMR parameters in monoclinic, insulating, $VO_2$ was performed using the experimental XRD structure (ICSD entry 34033[14]). The structure was relaxed using a Hubbard U correction of 0.5 eV to account for the electron correlations and stabilise the insulating structure, then the NMR parameters were calculated without the Hubbard U but with fixed electronic occupancy so as to retain the insulating ground state. Table 1 shows the calculated $^{17}$O chemical shifts, converted from shielding values using the regression of Figure 2; although the exact agreement with the experimental values is not excellent, the difference between the two oxygen environments is clear, enabling their assignment. Shown also are the quadrupolar coupling constants of the two environments, which were converted to the second-order quadrupolar shifts expected at 7.05 T using the MagresView software package.[15]

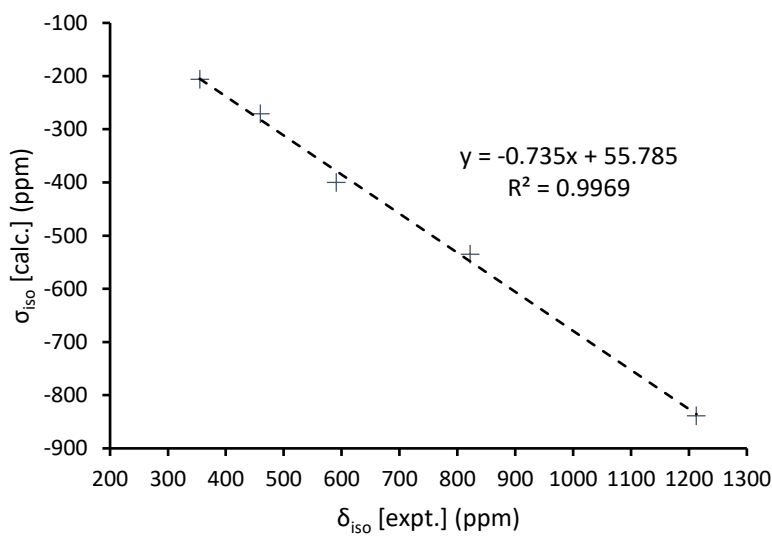

*Figure 2: Calculated $^{17}$O NMR isotropic shielding values plotted against the experimental isotropic chemical shift for selected diamagnetic binary first-row transition metal oxides.*

*Table 1: Calculated and experimental $^{17}$O chemical shifts in monoclinic, insulating, $VO_2$ as well as the calculated quadrupolar coupling constants, $C_Q$, and resultant second-order quadrupolar shifts at 7.05 T.*

|  | Experimental Shift /ppm | Calculated Shift /ppm | Calculated $C_Q$ /MHz | Second order quadrupolar shift /ppm |
|---|---|---|---|---|
| **Inter-dimer** | 753 | 696 | 1.72 | −12 |
| **Intra-dimer** | 814 | 832 | 1.56 | −10 |

## 3 Thermogravimetric Analysis

The hydrogen content of Pd/$H_x$VO$_2$ was determined from the mass change due to loss of H$_2$O on heating the sample to 600 °C in flowing N$_2$ using a Mettler Toledo TGA/SDTA 851 thermobalance with a 100 μL Al$_2$O$_3$ crucible (Figure 3), according to the following equation:

$$\mathrm{Pd}_{0.008}\mathrm{H}_x\mathrm{VO}_2 \rightarrow \mathrm{Pd}_{0.008}\mathrm{VO}_{\left(2-\frac{x}{2}\right)} + \frac{x}{2}\mathrm{H}_2\mathrm{O}$$

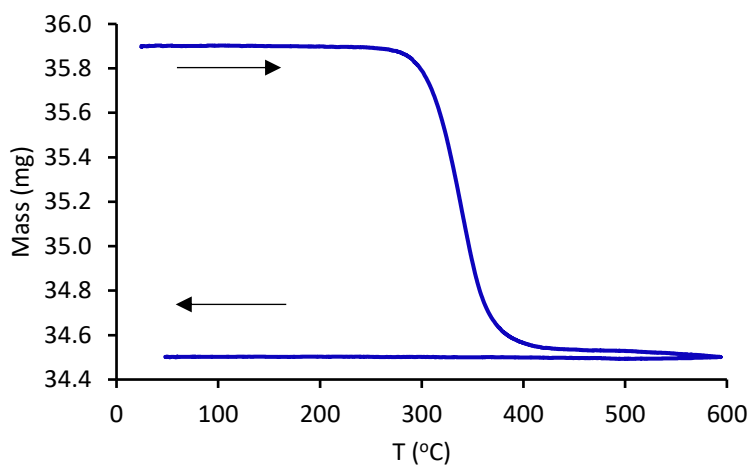

Figure 3: Mass of Pd/$H_x$VO$_2$, synthesised at 180 °C, upon heating to 600 °C under flowing N$_2$.

## 4 Rietveld Analysis

Figures 5 to 11 show the powder XRD patterns of selected $H_x$VO$_2$ samples, the identified phases and the Rietveld refinements. The purpose of the analysis is to identify the phases present and quantify the phase fractions, but the data is of insufficient quality for a full refinement of the atomic positions and displacement parameters and this was not attempted; these are the most likely causes of the discrepancies between the Rietveld refinements and the data. The atomic positions were fixed as shown in Table 2.

Table 2: Atomic positions in VO$_2$ and $H_x$VO$_2$.

| Phase | Atom | x | y | z |
|---|---|---|---|---|
| VO$_2$ M | V | 0.240 | 0.982 | 0.032 |
| | O | 0.106 | 0.21 | 0.203 |
| | O | 0.416 | 0.735 | 0.316 |
| $H_x$VO$_2$ O1a & O1b | V | 0 | 0 | 0 |
| | O | 0.279 | 0.311 | 0 |

The zero error was determined for the films from the PTFE reflection at 2θ = 18.21°. A preferred orientation along the [110] direction was observed for the orthorhombic phases and corrected for; although the preferred orientation effect could be removed for the Pd/$H_x$VO$_2$ sample by sprinkling the powder on vacuum grease, the signal intensity was then much worse, and this was also not possible for the PTFE films. The lattice parameters and phase fractions were then refined (Table 3).

To determine the hydrogen content of the orthorhombic phases from the cell parameters, the relationship between the orthorhombic distortion and the hydrogen content was determined from the data of Chippindale et al.[16] (Figure 4). The total hydrogen content of the sample could then be predicted by multiplying the hydrogen content of each phase by the phase fraction; the contribution from the monoclinic phase, if present, cannot be reliably determined from the cell parameters, so is not included in this approach.

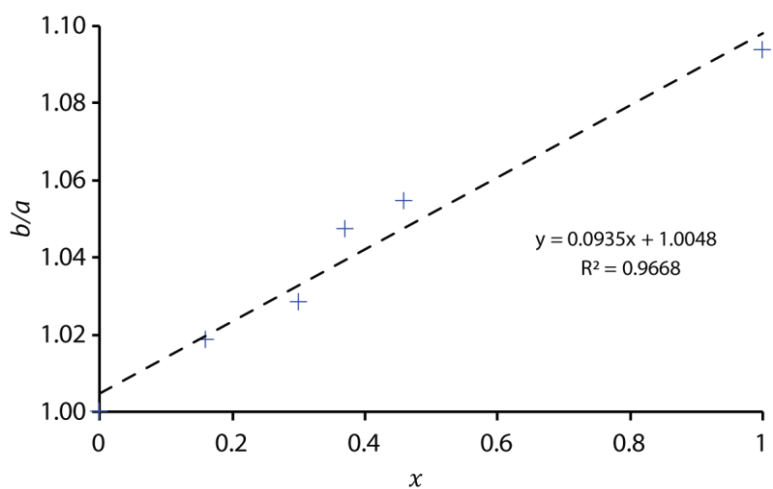

Figure 4: The orthorhombic distortion, b/a, plotted against the hydrogen content, $x$ in $H_xVO_2$, for the data of Chippindale et al.[16]

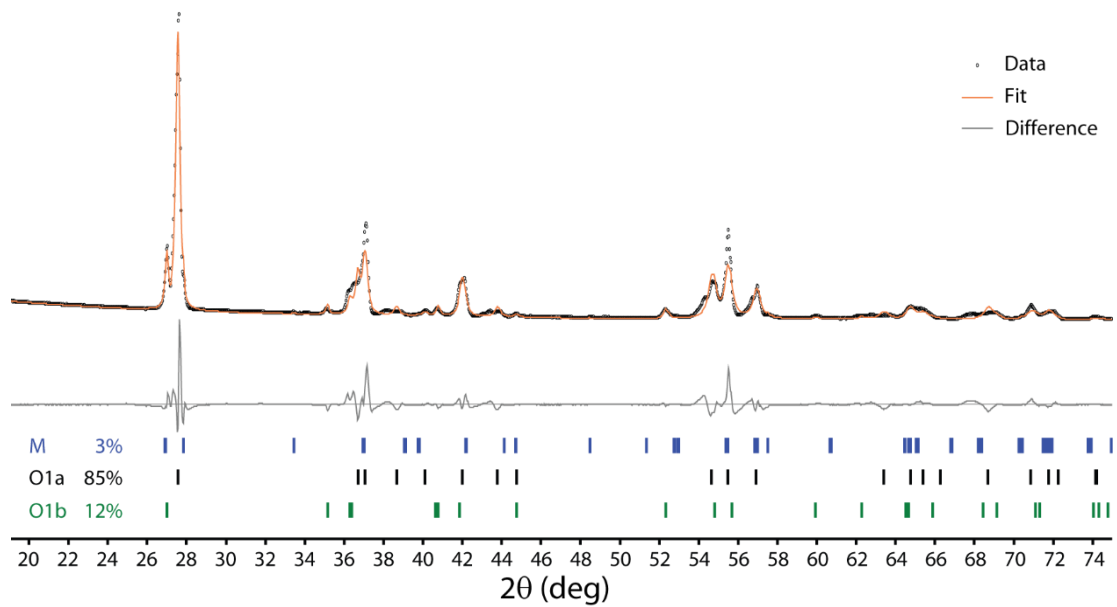

Figure 5: XRD pattern and Rietveld refinement of $VO_2$ catalytically hydrogenated at 180 °C with 25% $H_2/N_2$.

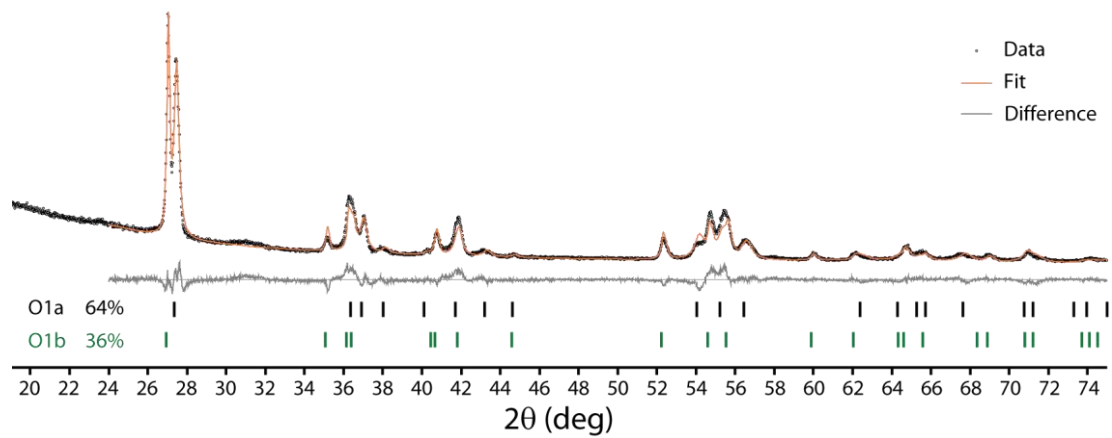

*Figure 6: XRD pattern and Rietveld refinement of VO$_2$ catalytically hydrogenated at 220 °C with 5% H$_2$/Ar.*

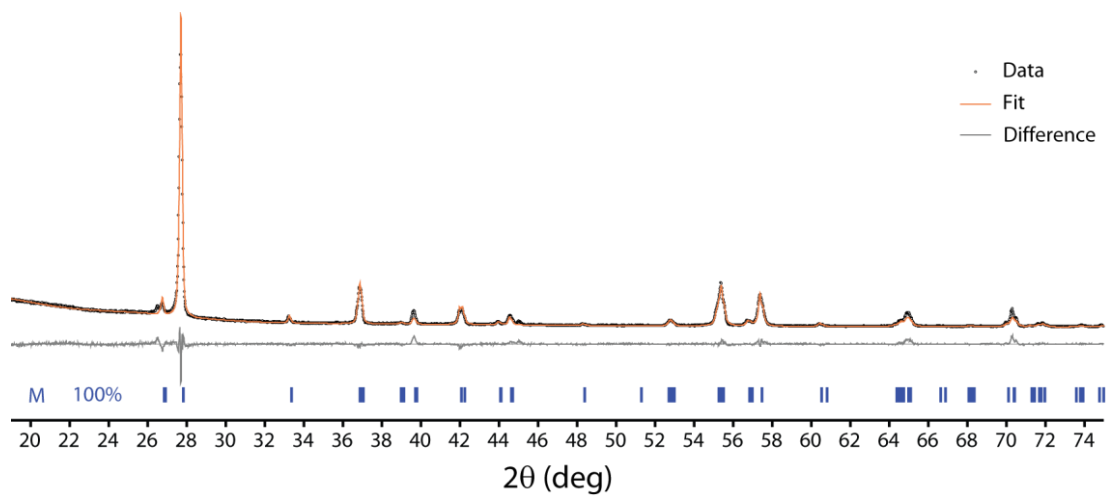

*Figure 7: XRD pattern and Rietveld refinement of VO$_2$ electrochemically hydrogenated at 21 °C.*

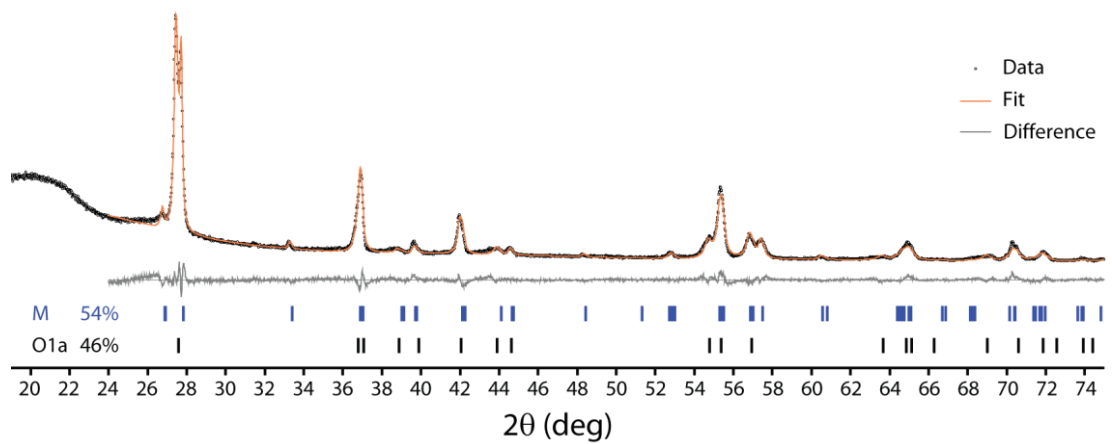

*Figure 8: XRD pattern and Rietveld refinement of VO$_2$ electrochemically hydrogenated at 50 °C.*

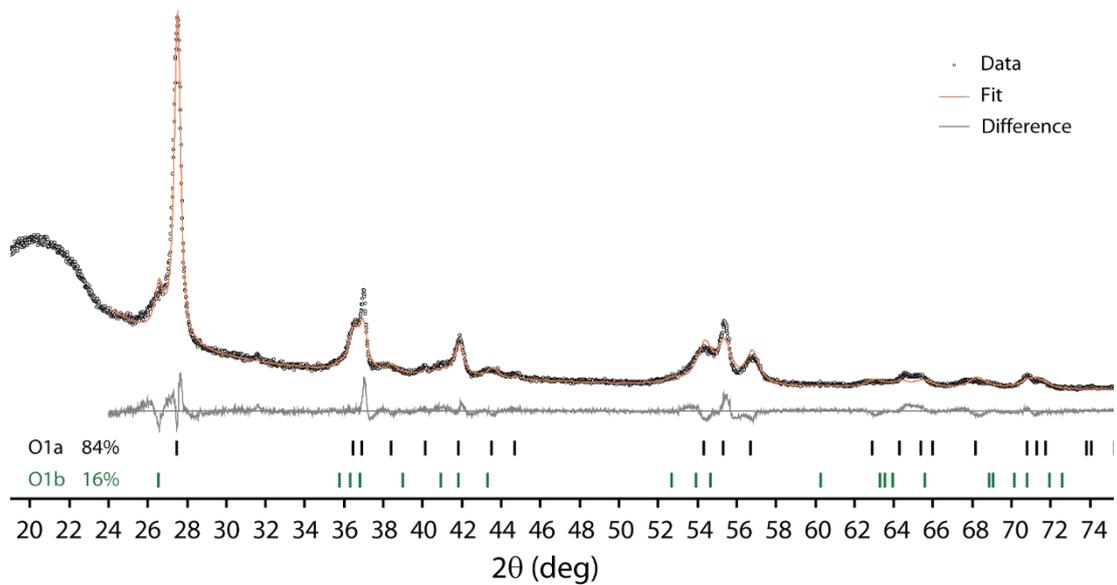

*Figure 9: XRD pattern and Rietveld refinement of VO$_2$ electrochemically hydrogenated at 100 °C*

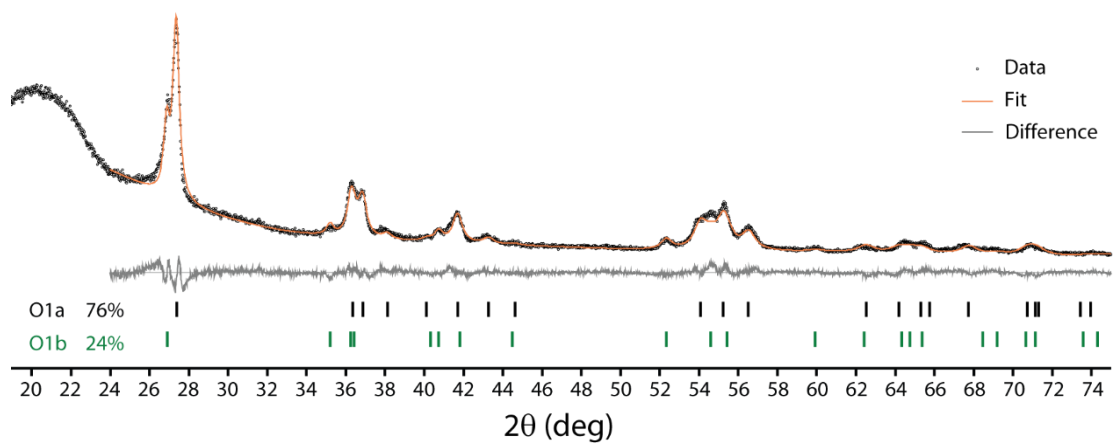

*Figure 10: XRD pattern and Rietveld refinement of VO$_2$ electrochemically hydrogenated at 150 °C.*

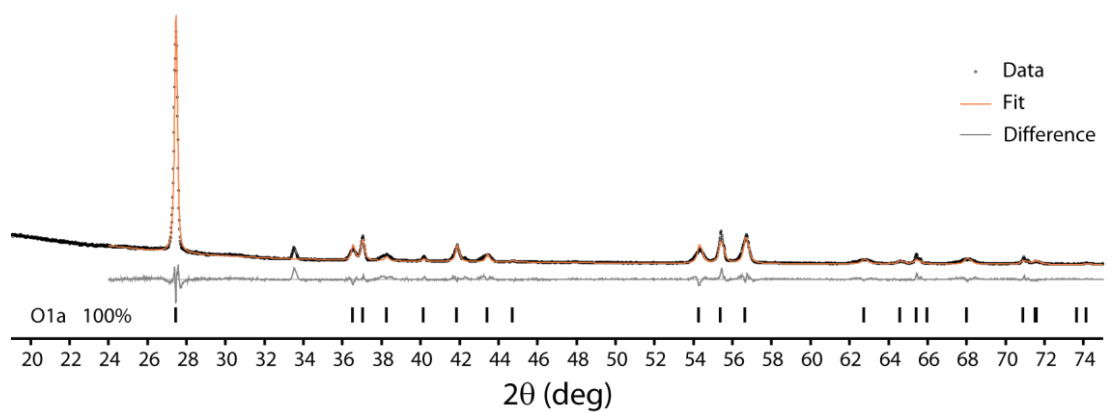

*Figure 11: XRD pattern and Rietveld refinement of VO$_2$ electrochemically hydrogenated at 200 °C. The peak at 2θ = 33.5° is ascribed to an unknown impurity or minor product.*

Table 3: Phase fractions and lattice parameters of the monoclinic and two orthorhombic phases in different $H_xVO_2$ samples, as well as the hydrogen content, $x$, of the phase as determined from the orthorhombic distortion. The contribution of the phase to the total hydrogen content of the sample, $x_{eff}$, is found by multiplying the hydrogen content of the phase by its phase fraction, and the total hydrogen content of the sample, $x_{tot}$, is then predicted by summing the contributions from each phase.

|  |  | Pristine $VO_2$ | Catalytic Hydrogenation | | Electrochemical Hydrogenation | | | | |
|---|---|---|---|---|---|---|---|---|---|
|  |  |  | 180 °C 25% $H_2$ | 220 °C 5% $H_2$ | 21 °C | 50 °C | 100 °C | 150 °C | 200 °C |
| **M** | wt% | 100% | 2.6% |  | 100% | 54.1% |  |  |  |
|  | $a$ | 5.753 | 5.753 |  | 5.756 | 5.755 |  |  |  |
|  | $b$ | 4.526 | 4.526 |  | 4.527 | 4.526 |  |  |  |
|  | $c$ | 5.383 | 5.383 |  | 5.384 | 5.380 |  |  |  |
|  | $\beta$ | 122.6 | 122.6 |  | 122.6 | 122.6 |  |  |  |
| **O1a** | wt% |  | 85.5% | 64.2% |  | 45.9% | 83.5% | 75.6% | 100.0% |
|  | $a$ |  | 4.494 | 4.494 |  | 4.515 | 4.491 | 4.494 | 4.489 |
|  | $b$ |  | 4.654 | 4.729 |  | 4.631 | 4.688 | 4.718 | 4.704 |
|  | $c$ |  | 2.877 | 2.895 |  | 2.874 | 2.895 | 2.899 | 2.885 |
|  | $x$ |  | 0.329 | 0.508 |  | 0.224 | 0.417 | 0.484 | 0.462 |
|  | $x_{eff}$ |  | 0.280 | 0.326 |  | 0.103 | 0.348 | 0.366 | 0.462 |
| **O1b** | wt% |  | 11.9% | 35.8% |  |  | 16.5% | 24.4% |  |
|  | $a$ |  | 4.444 | 4.455 |  |  | 4.545 | 4.470 |  |
|  | $b$ |  | 4.941 | 4.935 |  |  | 4.944 | 4.930 |  |
|  | $c$ |  | 2.981 | 2.990 |  |  | 2.895 | 2.974 |  |
|  | $x$ |  | 1.144 | 1.103 |  |  | 0.889 | 1.051 |  |
|  | $x_{eff}$ |  | 0.153 | 0.395 |  |  | 0.147 | 0.256 |  |
| **Total** | $x_{tot}$ |  | 0.433 | 0.721 |  | 0.103 | 0.495 | 0.622 | 0.462 |

Figure 12 shows a comparison of the hydrogen contents determined by quantitative $^1$H NMR and Rietveld analysis of the powder XRD patterns. The positive correlation supports the assignment of the lower and higher shift regions of the $^1$H NMR spectrum to the O1a and O1b phases respectively, and the use of the orthorhombic distortion to predict the hydrogen content from the unit cell parameters. However, there is also both scatter in the values greater than the predicted uncertainty as well as systematic deviation. The scatter is most likely due the difficulty in accurately determining the phase fractions from Rietveld analysis with broad peaks, as small changes in amplitude correspond to large changes in mass fraction. The quantitative $^1$H NMR also generally predicts a lower hydrogen content than the XRD analysis; this is most likely due to NMR skin depth effects: the metallic nature of the samples reduces radiofrequency penetration so that the sample is not fully excited and the $^1$H NMR signal is reduced. This was be circumvented for the catalytically hydrogenated Pd/$H_x VO_2$ samples by grinding with KBr, and for these samples, the NMR and XRD determined hydrogen contents are in better agreement. However, for the electrochemically hydrogenated $VO_2$, the PTFE films could not be ground with KBr, and although the presence of PTFE reduces the skin depth effect, there is still a notable contribution. It can also be seen that the skin depth effect mostly contributes to the metallic O1a phase, rather than the paramagnetic O1b phase.

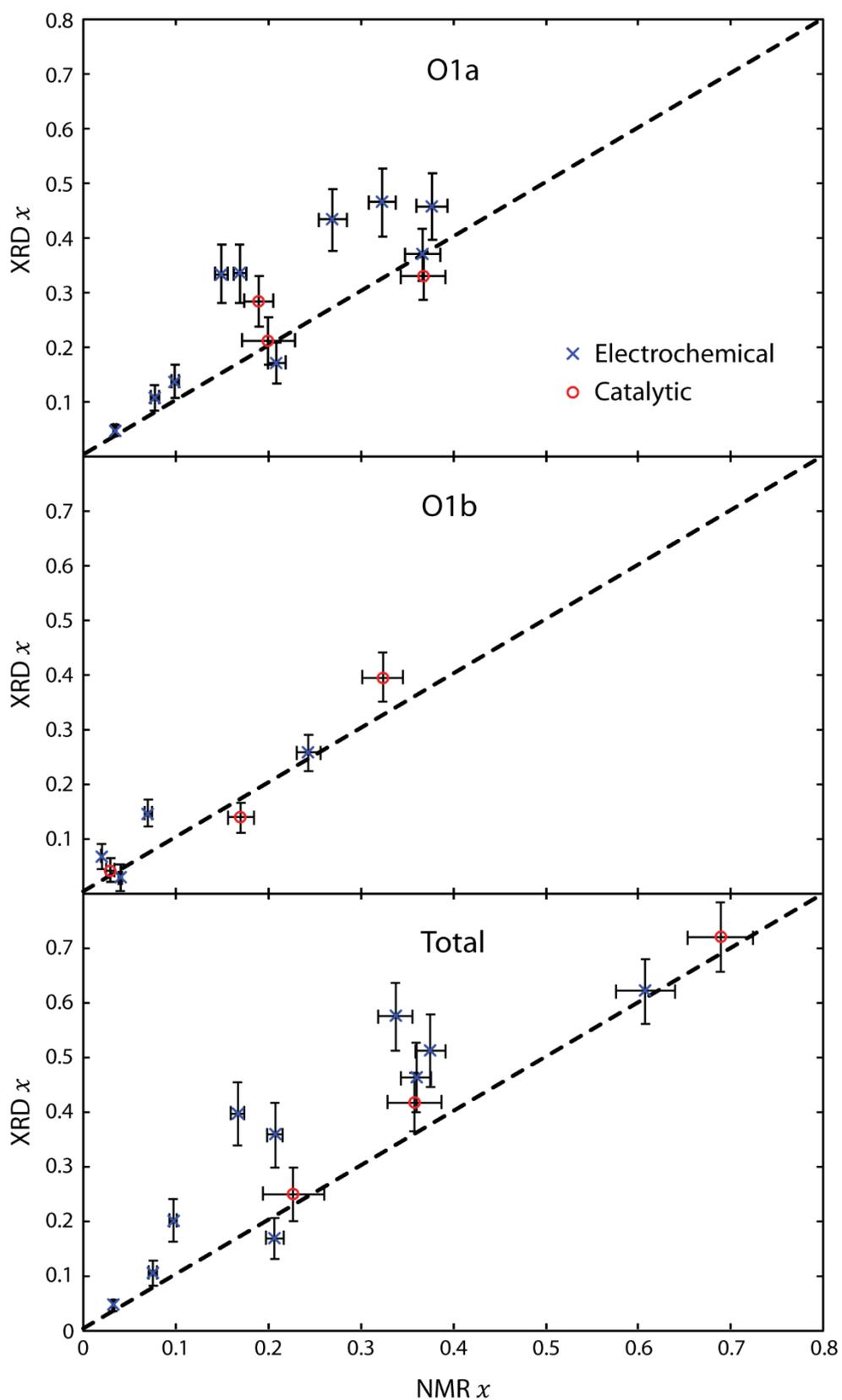

*Figure 12: Comparison of the hydrogen content of each orthorhombic phase and the total hydrogen content for various $H_xVO_2$ samples as determined by quantitative $^1H$ NMR and by the unit cells and phase fractions from Rietveld refinement of the powder XRD patterns. The dashed line indicates 1:1 agreement between the methods.*

# 5 ¹H NMR of Different Catalytically Hydrogenated Samples

Two samples of VO$_2$ were catalytically hydrogenated, one using 25% H$_2$/N$_2$ at 180 °C and one using 5% H$_2$/Ar at 220 °C. Rietveld analysis of the former gave a ratio of the two orthorhombic phases O1a:O1b of 85:12 compared to 64:36 for the latter (Figure 5, Figure 6 and Table 3). The ¹H NMR spectra of the two samples are shown in Figure 13; the second sample has a much higher ratio of the signal at 450 ppm to that at 100 ppm, suggesting that the two shift regions correspond to the O1b and O1a phases respectively. This assignment is supported by the quantitative analysis of many samples, both electrochemically and catalytically hydrogenated (Figure 12).

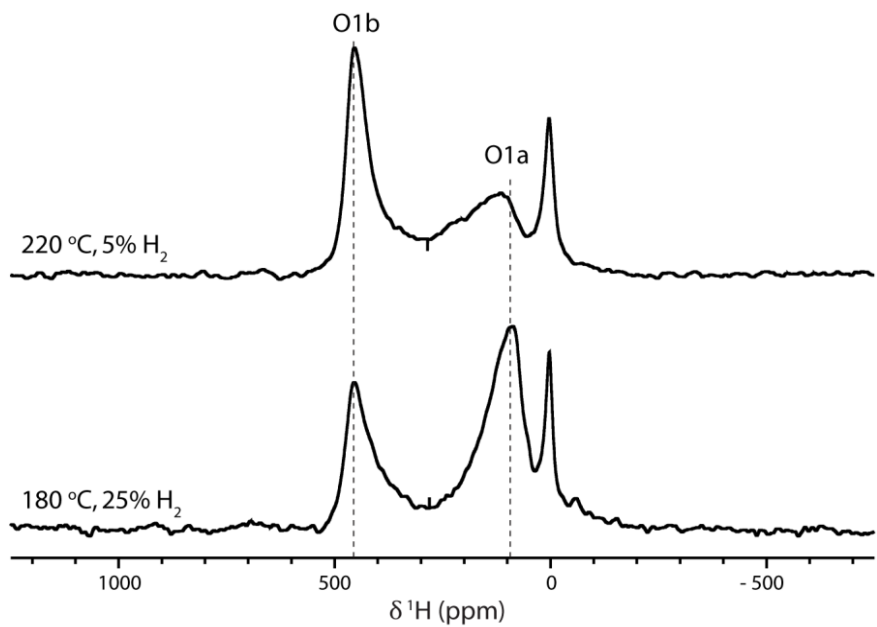

*Figure 13: ¹H NMR spectra of catalytically hydrogenated VO$_2$ samples prepared* via *two different procedures. The spectra were recorded at 4.70 T and 40 kHz MAS using a MATPASS sideband separation pulse sequence and taking the isotropic slice.*

# 6 Variable Temperature ¹H NMR

Figure 14 shows the ¹H MATPASS NMR spectra of Pd/H$_x$VO$_2$ between 18 °C and 69 °C. The sample was ground with KBr so that the sample temperature could be determined *in-situ* by the temperature-dependent KBr $T_1$ constant and chemical shift.[17] The O1a peak appears to move to higher frequency with increasing temperature, but this is actually due to the lower frequency regions relaxing more quickly at higher temperatures and hence less of this signal being observed relative to the higher frequency regions; this effect makes it challenging to determine whether the chemical shifts are temperature-dependent, but the chemical shift certainly does not decrease with increasing temperature, ruling out a paramagnetic shift. The O1b shift, on the other hand, does decrease with increasing temperature; to determine whether this shift followed the expected Curie–Weiss temperature dependence, a wider temperature range was explored using a 4 mm rotor with a zirconia cap, again with an *in-situ* KBr thermometer. A single pulse experiment was used because the lower maximum spinning speed of 14 kHz, for the larger rotor, and fast T$_2$ relaxation of the sample prevented rotor synchronised experiments such as a MATPASS or a Hahn echo; the background in the single pulse experiment then obscured the O1a signal so that only the O1b signal could be distinguished. Figure 15a shows a plot of 1/shift versus temperature for the O1b signal, which shows the expected linear Curie–Weiss dependence, confirming that this is a paramagnetic shift, which is positive due to the 90° π delocalisation pathway (Figure 15b).

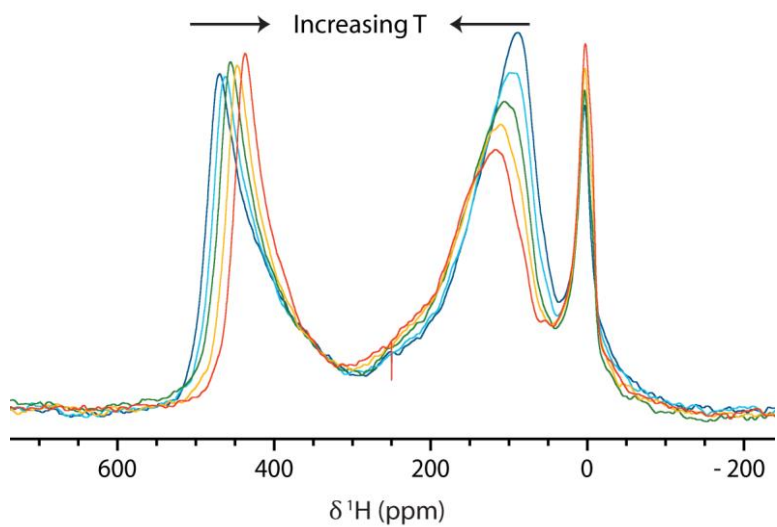

*Figure 14: Variable temperature $^1$H MATPASS NMR (7.05 T) spectra at 60 kHz MAS with sample temperatures between 18 °C and 69 °C as determined from the in-situ KBr NMR thermometer.*

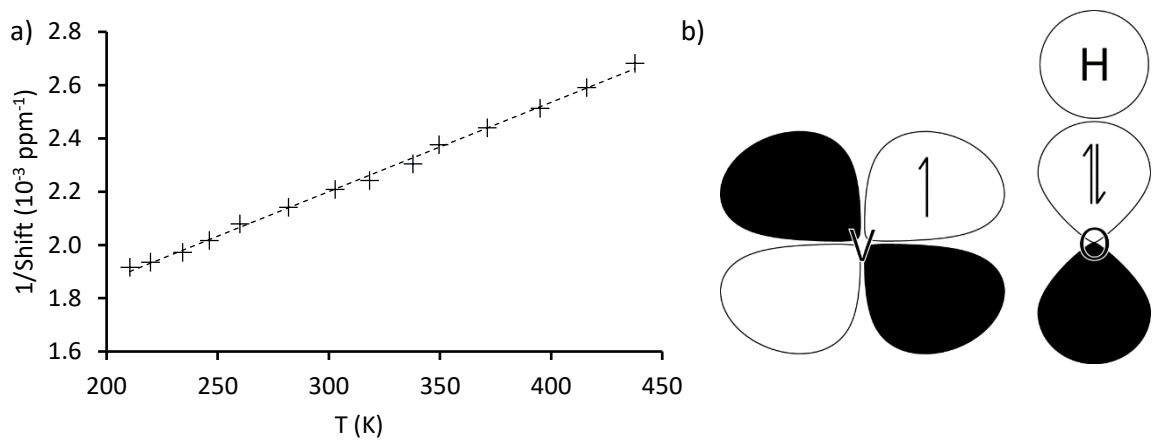

*Figure 15: a) A plot of 1/shift against T for the $^1$H O1b shift in Pd/H$_x$VO$_2$. The spectra were recorded using a single-pulse experiment at 14 kHz MAS and 4.70 T. The sample temperature was determined with an in-situ KBr NMR thermometer. b) Schematic representation of the 90° π delocalisation pathway which gives a positive paramagnetic shift on the proton.*

# 7   $^1$H $T_1$ measurements

To measure the $T_1$ relaxation constant for the $^1$H environments in H$_x$VO$_2$, an inversion recovery experiment was used. Both the O1a and O1b NMR signals represent a distribution of environments, but this can be modelled reasonably well by deconvoluting the signal from each phase as two Gaussian-Lorentzian peaks, as shown in Figure 16 for the MATPASS $^1$H spectrum. This permits the contribution from each signal to be determined in the $^1$H inversion recovery spectra as a function of the relaxation time (Figure 17), which in turn can be fitted to find the $T_1$ relaxation constants (Table 4). The relaxation is fast for both phases due to the metallic and paramagnetic environments (the $T_1$ constants of diamagnetic protons in insulating environments are typically 1–10 s), with the paramagnetic relaxation of the O1b signals an order of magnitude faster than that of the O1a signals, further supporting the assignment of the O1b signal as paramagnetically shifted. Furthermore, within the signal of each phase, the component with a higher shift has a shorter relaxation constant; this is as expected because a higher shift corresponds to a more metallic or more paramagnetic environment respectively, for which the relaxation is consequently faster.

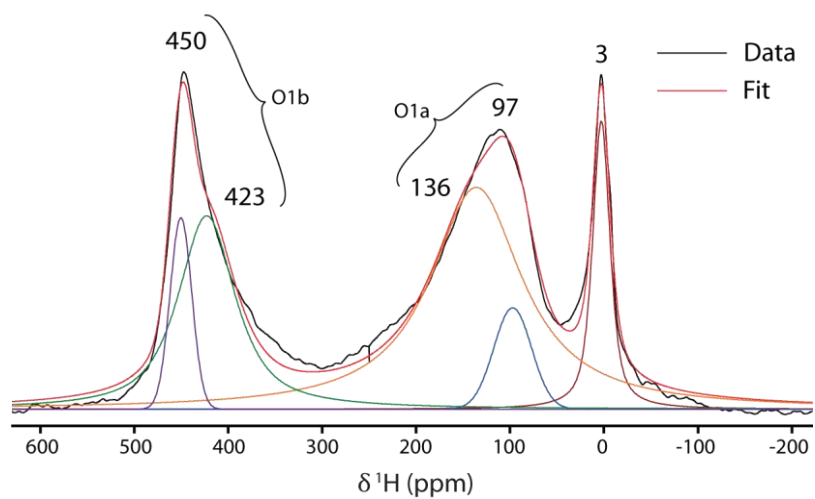

*Figure 16: Gaussian-Lorentzian deconvolution of the $^1$H MATPASS NMR spectrum of catalytically hydrogenated Pd/H$_x$VO$_2$, recorded at 7.05 T and 60 kHz MAS.*

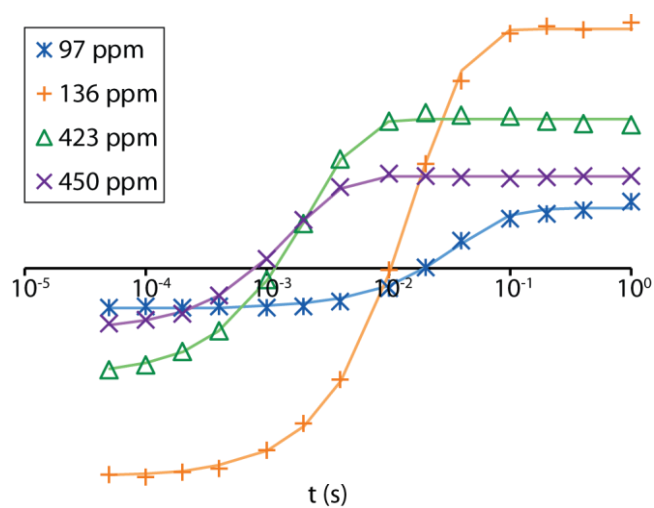

*Figure 17: Intensity as a function of relaxation time for different environments in the $^1$H inversion recovery NMR spectra of catalytically hydrogenated Pd/H$_x$VO$_2$.*

*Table 4: Fitted T$_1$ relaxation constants for different environments from the $^1$H inversion recovery NMR spectra of catalytically hydrogenated Pd/H$_x$VO$_2$.*

|         | 97 ppm | 136 ppm | 423 ppm | 450 ppm |
|---------|--------|---------|---------|---------|
| $T_1$/s | 0.039  | 0.017   | 0.0022  | 0.0016  |

# 8  Magnetisation *vs.* Field Measurements

Figure 18 shows the magnetisation *vs.* field measurements for VO$_2$ electrochemically hydrogenated at room temperature, recorded below (60 K) and above (300 K) the spin glass freezing temperature (T$_f$ ≈ 150 K). Below T$_f$ there is hysteresis in the magnetisation about the origin, which is no longer present above T$_f$; this is characteristic of a spin glass-like state.[18]

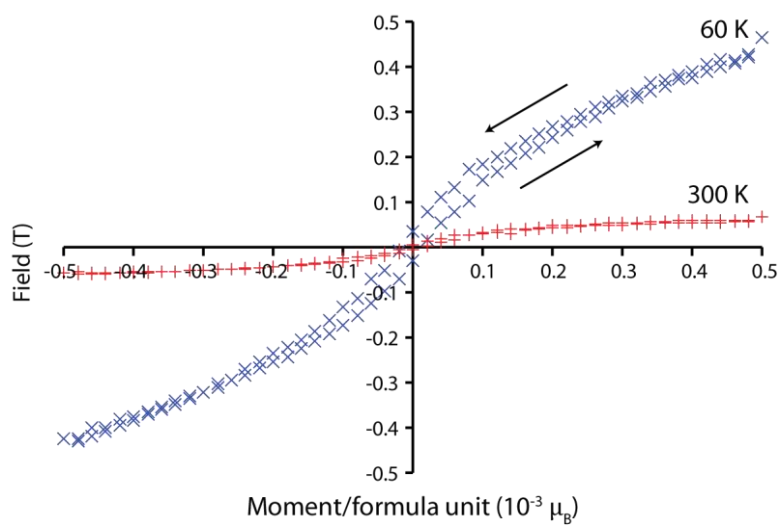

*Figure 18: M(H) curves for VO₂ electrochemically hydrogenated at room temperature, recorded below (60 K) and above (300 K) the spin glass freezing temperature ($T_f$).*

# 9 Comparison of the $^{17}$O NMR of Pristine and Electrochemically Hydrogenated VO$_2$

Figure 19 shows the $^{17}$O NMR spectra of $^{17}$O enriched VO$_2$ recorded above the MIT, before and after electrochemical hydrogenation with EMIm TFSI ionic liquid at room temperature. The two peaks from residual monoclinic insulating VO$_2$, present due to the finite width of the MIT and temperature gradients in the rotor, are at the same shift for both samples. However, the signal from metallic rutile VO$_2$ is shifted to more negative frequency for the electrochemically hydrogenated samples; this is evidence of the electron doping associated with hydrogen intercalation, which increases the density of states at the Fermi level, and hence the Knight shift.

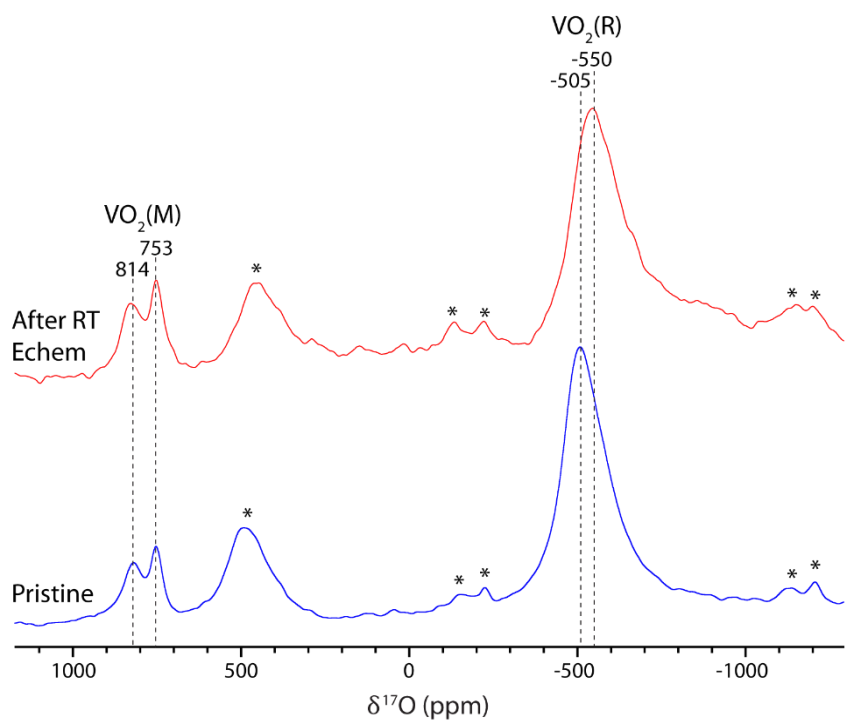

*Figure 19: $^{17}$O NMR spectra of $^{17}$O-enriched VO$_2$ above the MIT, before and after electrochemical hydrogenation with EMIm TFSI at room temperature, showing metallic rutile VO₂(R) and residual insulating monoclinic VO₂(M). Spectra were recorded at 69 °C, 7.05 T and 40 kHz MAS using a Hahn echo pulse sequence; sidebands are marked with an asterisk.*

## 10 Two-Electrode Potentiostatic Electrochemical Hydrogenation

Figure 20a shows the current and specific charge profiles for bulk VO$_2$ potentiostatically hydrogenated by applying −2.5 V between the Pt counter electrode and VO$_2$ working electrode for 11 hours, while Figure 20b shows the subsequent $^1$H NMR spectrum. The H$_x$VO$_2$ signal is clearly observed in the $^1$H NMR, indicating that hydrogenation has occurred, while quantification yields $x$ = 0.016, which corresponds to a specific charge of 5.2 mAh g$^{-1}$. The current rapidly decays then approaches a non-zero value of ~0.01 mA; the charge required for hydrogenation has been transferred within the first 30 minutes, so the subsequent and limiting current correspond to electrolyte breakdown, which is still appreciable even with an applied voltage of 2.5 V.

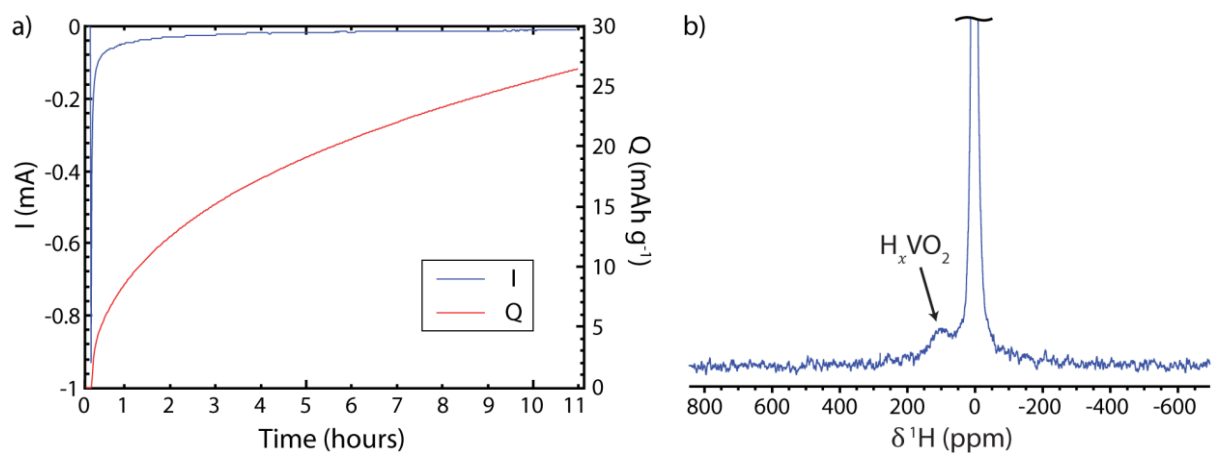

*Figure 20: a) Current and specific charge profiles and b) $^1$H NMR spectrum of potentiostatically hydrogenated VO$_2$. The spectrum was recorded at 40 kHz MAS and 4.70 T, taking the isotropic slice from a MATPASS spectrum.*

## 11 Reversible Electrochemical Hydrogenation

To test the reversibility of electrochemical hydrogenation, a bulk VO$_2$ sample was galvanostatically reduced for 24 hours at room temperature with EMIm TFSI, before reversing the current for 24 hours (Figure 21a). The $^1$H spectrum for this sample (Figure 21b) shows only the diamagnetic signal and the broad probe background, but no H$_x$VO$_2$ signal, indicating that the hydrogenation has been reversed. Closer examination of the voltage profile shows that during reduction the potential of the VO$_2$ working electrode reaches a limiting plateau, with only the initial sloping region corresponding to hydrogenation of VO$_2$. When the current is switched to VO$_2$ oxidation, the working electrode potential increases again while the VO$_2$ dehydrogenates, before reaching a limiting potential corresponding to the anodic limit of the ionic liquid – at approximately the same potential observed at the Pt counter electrode during reduction. Notably, however, on oxidation the potential of the Pt counter electrode is significantly lower than observed at the VO$_2$ electrode during its reduction: this indicates that the cathodic stability of the ionic liquid is decreased at a VO$_2$ electrode relative to a Pt electrode, possibly explaining why electrolyte breakdown reactions are significant in this system at voltages which are within the typically reported stability window.

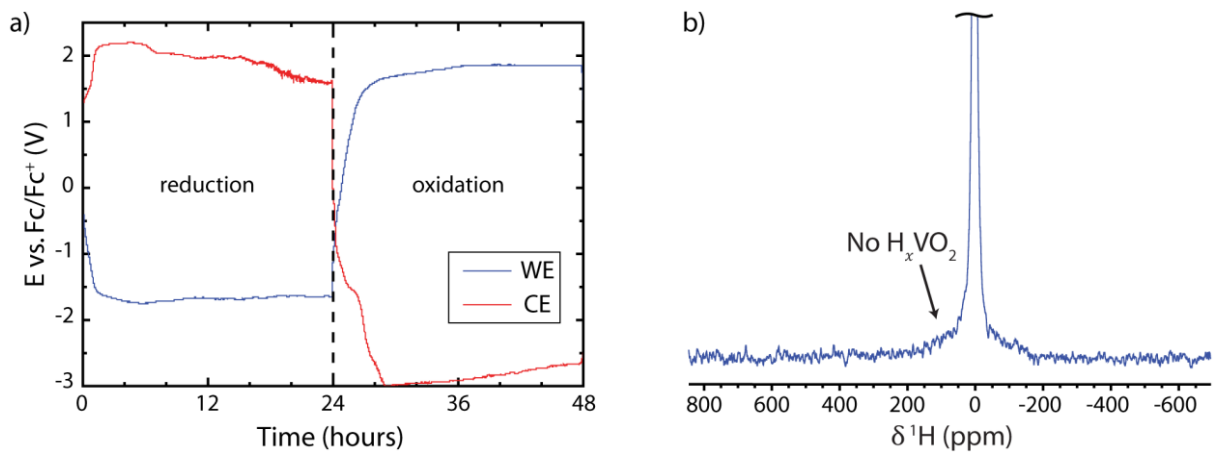

*Figure 21: a) Voltage profiles for the VO$_2$ working electrode (WE) and Pt counter electrode (CE), and b) $^1$H NMR spectrum of reversibly electrochemically hydrogenated VO$_2$. The spectrum was recorded at 40 kHz MAS and 4.70 T with a MATPASS pulse sequence.*

## 12 SEM of Bulk VO$_2$

Figure 22 shows a representative SEM micrograph of a bulk VO$_2$ film, made with 10 wt% PTFE binder and 10 wt% conductive carbon nanoparticles. Analysis of 100 particles using ImageJ gives an average particle size of 1.9 µm, with a standard deviation of 0.9 µm.

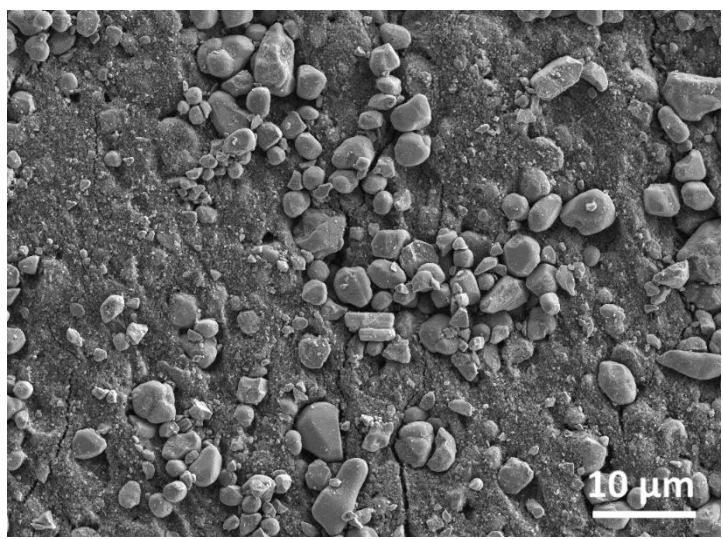

*Figure 22: SEM micrograph of bulk VO$_2$ with 10 wt% PTFE and 10 wt% conductive carbon nanoparticles.*

## 13 Electrochemical Hydrogenation of VO$_2$ Nanoparticles

To determine whether electrochemical hydrogenation of bulk VO$_2$ at room temperature was limited by the particle size, VO$_2$ nanoparticles were synthesised by ball milling. The comproportionated VO$_2$ (500 mg) was ball milled for 8 × 15 minutes in a 50 mL zirconia jar with five 10 mm diameter zirconia balls, using a Fritsch Pulverisette 23 shaker mill. The resultant nanoparticles were then characterised *via* powder XRD, SEM and scanning transmission electron microscopy (STEM). Analysis of the XRD peak widths yielded an average crystallite size of 12 nm, however this is an underestimate as it ignores any broadening due to strain. The electron microscopy (Figure 23) reveals micron-sized secondary particles, comprising primary particles averaging 30 nm in diameter, with particles as small as 10 nm being distinguishable in the STEM image.

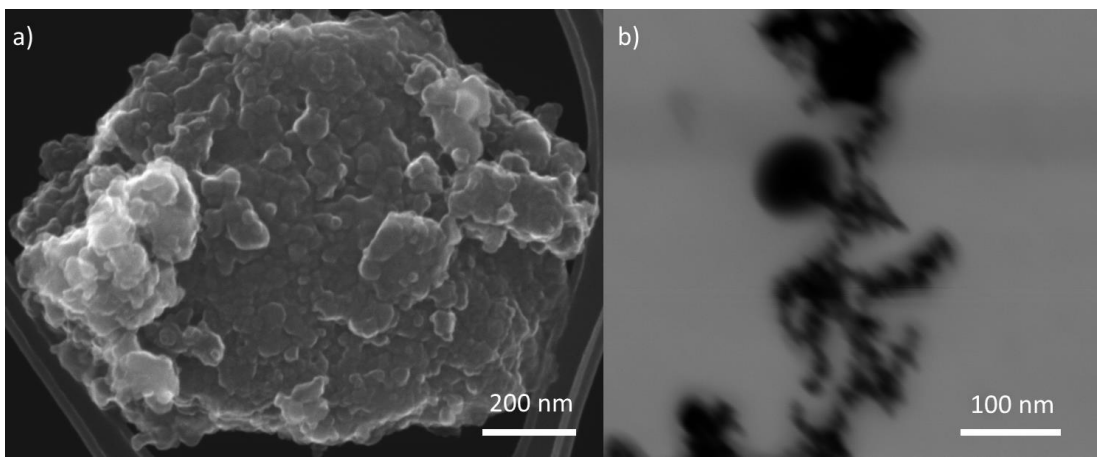

*Figure 23: a) SEM image of ball-milled VO$_2$ nanoparticles, showing a larger secondary particle comprising smaller primary particles, and b) STEM image showing some unagglomerated primary particles.*

From the VO$_2$ nanoparticles, a composite free-standing film was made with 10% PTFE and 10% conductive carbon. The film then underwent galvanostatic electrochemical reduction at room temperature (Figure 24). The voltage profile exhibits a less steep gradient than for bulk VO$_2$ and more charge is transferred before hitting the limiting plateau at around −1.6 V. The $^1$H NMR spectrum (Figure 25a) shows greater hydrogenation, and quantification yields a hydrogen content of $x$ = 0.20, *c.f.* 0.037 for bulk VO$_2$; the hydrogen content also matches well with the amount of charge transferred before reaching the limiting plateau. The $^{51}$V NMR spectrum (Figure 25b) does show a broad signal at negative shift due to vanadium in a metallic environment, however some of the signal from insulating VO$_2$(M) remains. Some side products are also evident, with the same diamagnetic vanadium signal as observed after electrochemical reduction of bulk VO$_2$ at 100 °C and 150 °C. The lack of uniform metallisation in this case is ascribed to the difficulty of electrically contacting all the nanoparticles with the conductive carbon, which is compounded by the presence of secondary agglomerates; it is likely that some particles are electrically or electrochemically isolated and hence are not reduced, giving rise to the residual VO$_2$(M) signal. Nevertheless, the greater hydrogenation observed for nanoparticulate VO$_2$ as compared to bulk VO$_2$ suggests that particle size is the reason full metallisation is not observed for the latter. This is most likely due to sluggish kinetics of hydrogen diffusion and/or nucleation and growth of the new phases, relative to competing electrochemical side reactions such as hydrogen evolution.

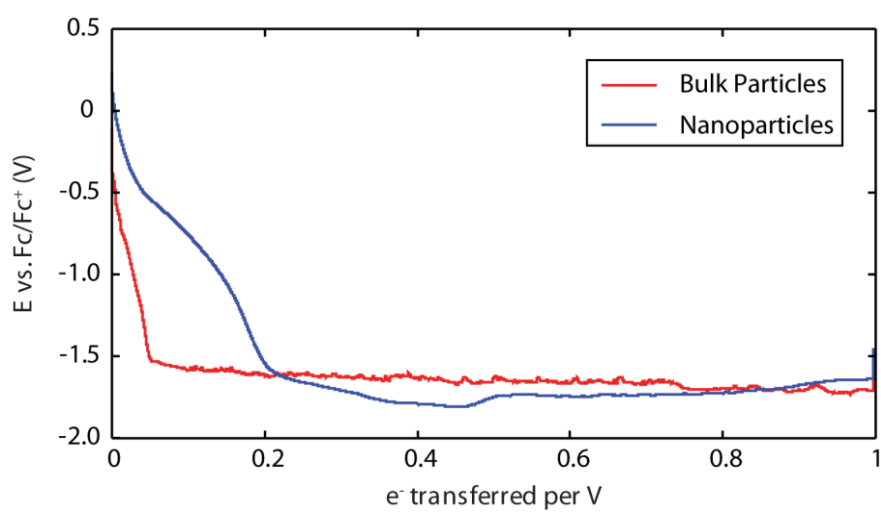

*Figure 24: Voltage profiles for bulk (~2 μm) and nanoparticulate (~30 nm) VO$_2$ electrochemically reduced at room temperature with EMIm TFSI.*

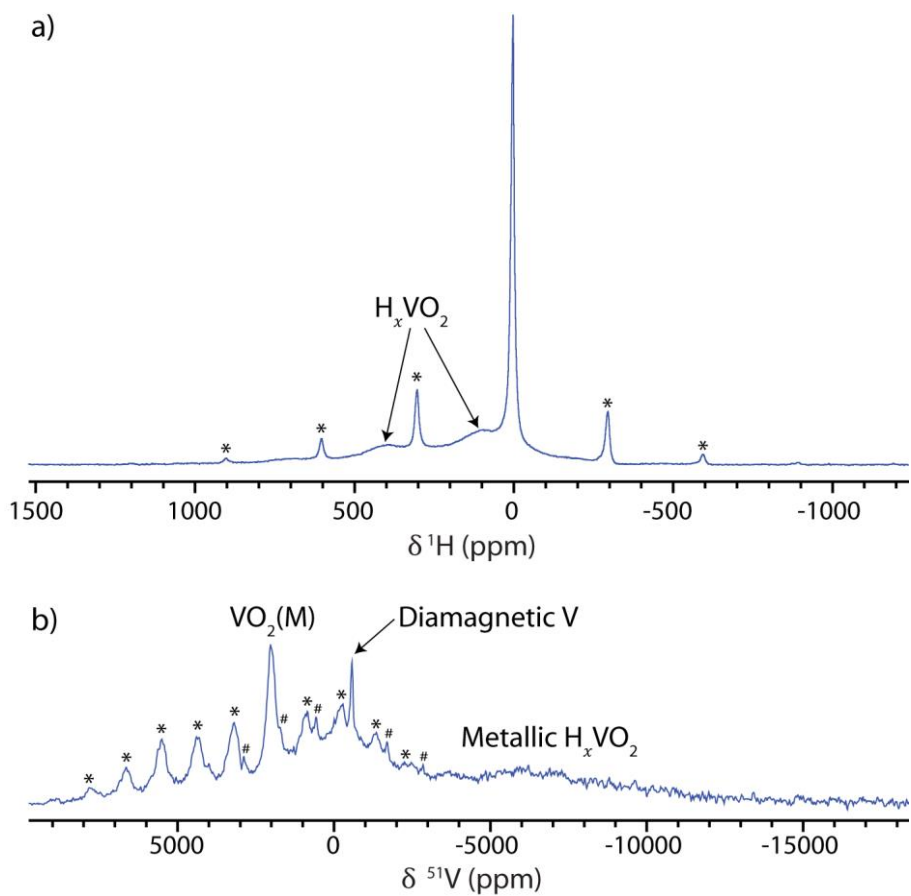

*Figure 25: a) $^1$H and b) $^{51}$V NMR spectra of electrochemically hydrogenated VO$_2$ nanoparticles, recorded at 4.70 T and 60 kHz MAS using a Hahn echo pulse sequence. The $^{51}$V spectrum was acquired in steps of 5000 ppm between carrier frequencies of 5000 ppm and −20000 ppm and summed to produce the spectrum shown above. Spinning sidebands are marked with asterisks or hash marks.*

## 14 Magnetic Susceptibility Measurements of H$_x$VO$_2$ Electrochemically Hydrogenated at Elevated Temperatures

Figure 26 shows the bulk magnetic susceptibility measurements for pristine VO$_2$ and H$_x$VO$_2$ which has been electrochemically metallised between 21 °C and 200 °C. The general trend is an increase in both the Pauli (high T asymptote) and Curie–Weiss (low T tail) paramagnetism with increasing temperature up to 150 °C, due to the increased hydrogenation (main text, Figure 5). The loss of the MIT can also be seen for electrochemical hydrogenation above 50 °C. A maximum in the susceptibility, corresponding to antiferromagnetic ordering, is observed for the samples metallised at the highest three temperatures, at Néel temperatures of ~8 K for the 100 °C and 150 °C samples and ~25 K for the 200 °C sample; see section 17 for a discussion of the 200 °C sample. The fit of the Curie–Weiss paramagnetic component from the main text used only points above the Néel temperature.

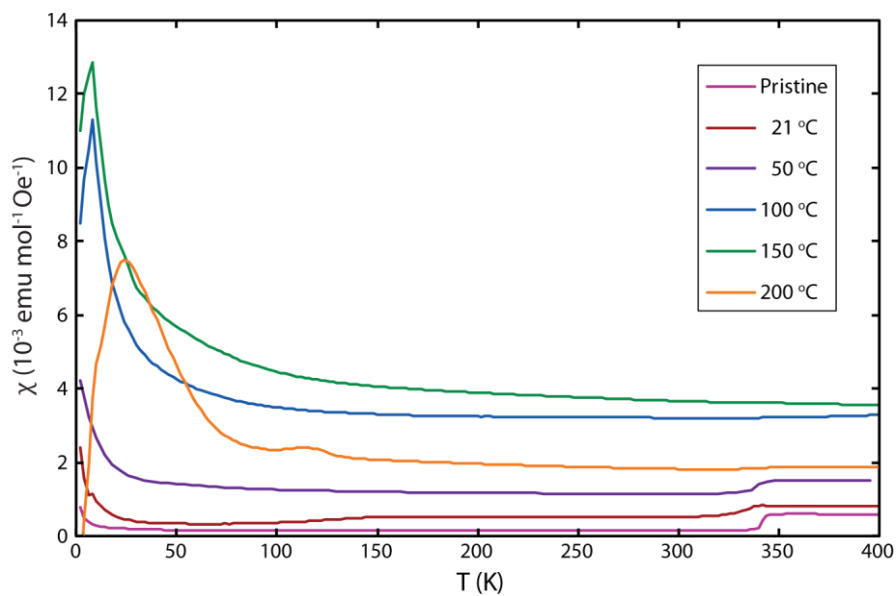

*Figure 26: ZFC magnetic susceptibility measurements of pristine VO$_2$ and H$_x$VO$_2$ which has been electrochemically metallised between 21 °C and 200 °C.*

## 15  $^{17}$O NMR Spectra of H$_x$VO$_2$ Electrochemically Hydrogenated at Elevated Temperatures

Figure 27 shows the $^{17}$O NMR spectra of H$_x$VO$_2$ electrochemically hydrogenated between 50 °C and 200 °C, recorded above and below the MIT temperature. In the low temperature spectrum, the sample which was electrochemically hydrogenated at 50 °C retains the insulating monoclinic VO$_2$ signals at 753 and 814 ppm, as was also observed for the room temperature sample (main text, Figure 4e). Then in the spectrum recorded above the MIT, the negatively Knight-shifted signal of the metallic phase is again observed at the more negative shift of −550 ppm, compared to −505 ppm for pristine VO$_2$; in the same way as for room temperature electrochemical hydrogenation, this is evidence of the electron doping associated with hydrogen intercalation, which increases the density of states at the Fermi level, and hence the Knight shift. Moreover, because this is the same shift as was observed for the room temperature sample, this suggests that the degree of hydrogenation of the monoclinic phase is the same for both samples; the 50 °C sample differs in that it also contains the more hydrogenated O1a phase.

For the sample electrochemically hydrogenated at 100 °C, the insulating monoclinic VO$_2$ signals are no longer present in the low temperature spectrum, nor is the sharp metallic signal observed in the high temperature spectrum; both spectra instead exhibit a very broad signal with unresolved sidebands, centred at ∼−1100 ppm; this can be more readily seen in the MATPASS spectrum (Figure 28, bottom). The shift of this signal is independent of temperature, so can be assigned as the metallic O1a phase with a core-polarisation negative Knight shift. Again, the shift is more negative due to an increase in the density of states at the Fermi level, due to greater doping of electrons, and the width of the signal may indicate a distribution of local environments or doping levels. The signal from the O1b phase cannot be distinguished, possibly because it is too broad; the peaks in the XRD pattern of VO$_2$ electrochemically metallised at 100 °C are also very broad (Figure 9), suggesting a distribution of lattice parameters or low crystallinity. Based on the appearance of O1a for this 100 °C sample, it is likely that the O1a phase present in the sample hydrogenated at 50 °C is responsible for the broad background observed in the $^{17}$O spectra of that sample, both above and below the MIT.

The sample electrochemically hydrogenated at 150 °C shows a sharper signal at ∼−1500 ppm as well as a shoulder at −1200 ppm; again, this can be more easily seen in the MATPASS spectrum (Figure 28, bottom). The sharp signal exhibits a Curie–Weiss temperature dependence (Figure 29a), so is assigned to the localised paramagnetic O1b phase; the paramagnetic shift is negative due to the polarisation mechanism of the orthogonal V $t_{2g}$ and O 2s orbitals,[19] and becomes less negative with increasing temperature. The shift of the shoulder is independent of temperature and is assigned to the O1a phase, once more with a more negative Knight shift. See Section 17 for a discussion of the 200 °C sample.

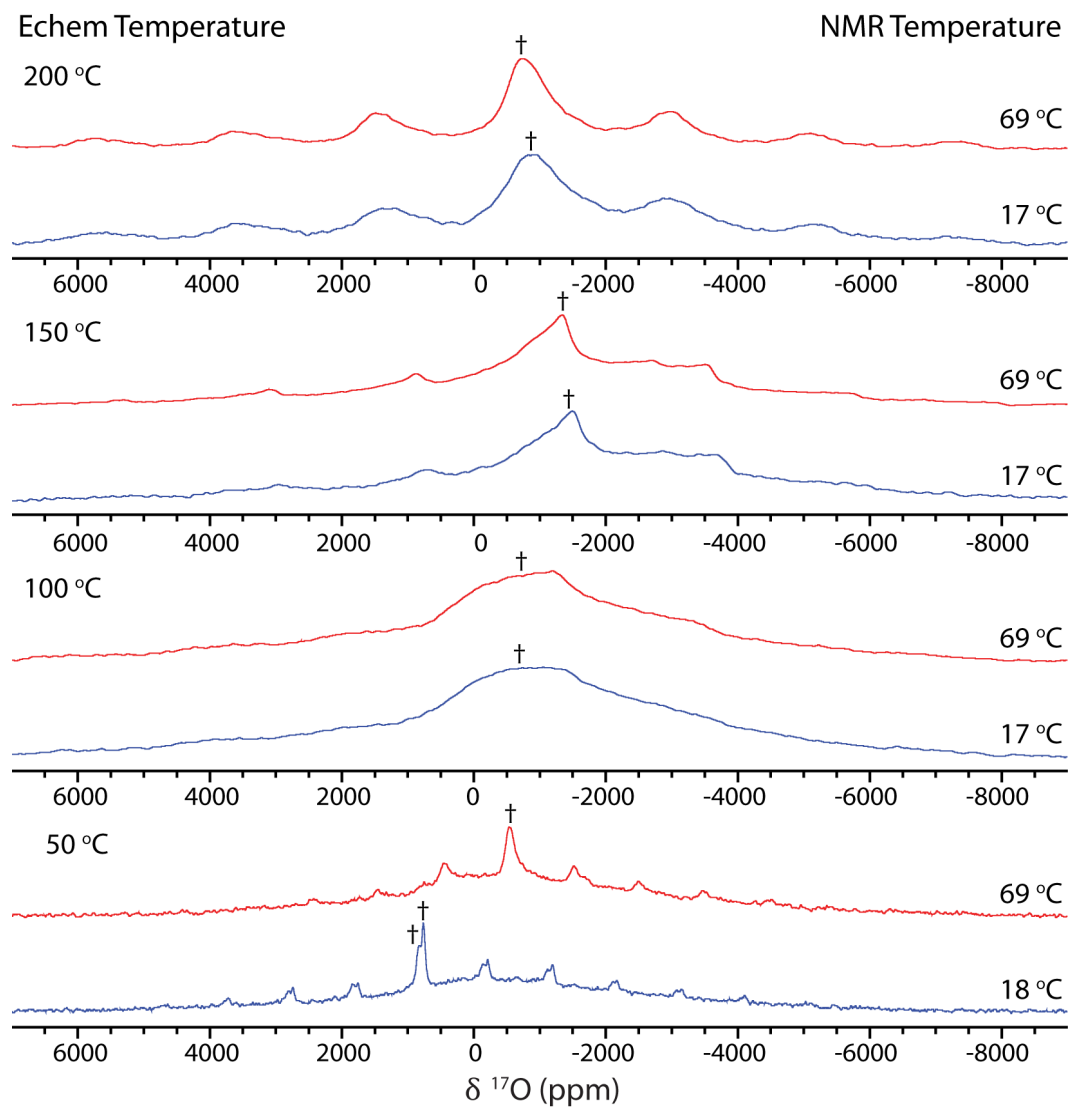

Figure 27: $^{17}$O NMR spectra of $^{17}$O-enriched H$_x$VO$_2$, electrochemically hydrogenated between 50 °C and 200 °C, recorded above and below the VO$_2$ MIT temperature, using a Hahn echo pulse sequence. The sample electrochemically hydrogenated at 50 °C was recorded at 7.05 T and 40 kHz MAS to afford better resolution of the monoclinic peaks, while the other samples were recorded at 4.70 T and 60 kHz MAS to achieve better sideband separation of the metallic and paramagnetic signals. The isotropic resonances are marked with a dagger.

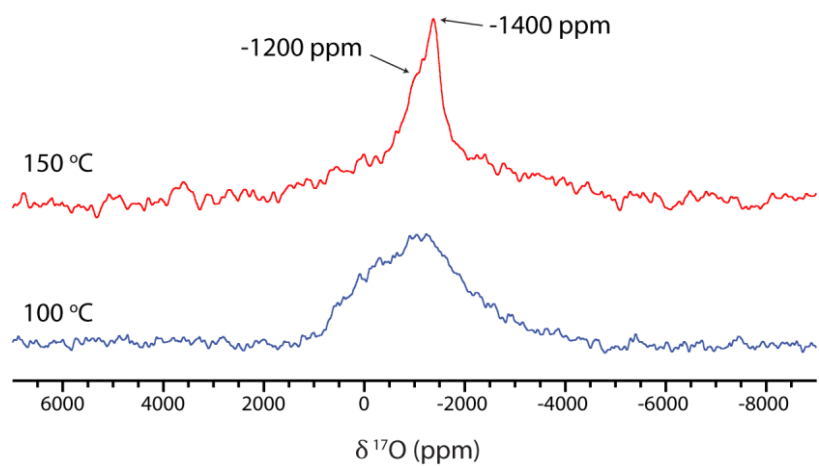

*Figure 28: The isotropic slices of the $^{17}O$ MATPASS NMR spectra of $^{17}O$-enriched $H_xVO_2$ electrochemically hydrogenated at 100 °C and 150 °C. The spectrum of the 100 °C sample was recorded at 7.05 T, 60 kHz MAS and 32 °C; the spectrum of the 150 °C sample was recorded at 4.70 T, 60 kHz MAS and 47 °C.*

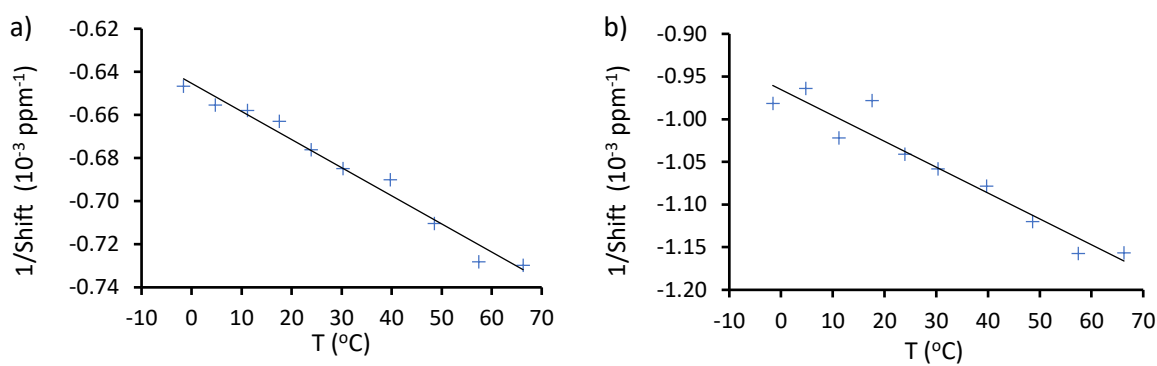

*Figure 29: A plot of 1/shift against T for the $^{17}O$ NMR spectra of $^{17}O$ enriched $H_xVO_2$ electrochemically gated at a) 150 °C and b) 200 °C. The spectra were recorded using a Hahn echo at 60 kHz MAS and 4.70 T. The sample temperature was determined by an ex-situ calibration with the $^{207}Pb$ signal of $Pb(NO_3)_2$.*

# 16 Voltage Profiles for Electrochemistry at Elevated Temperatures

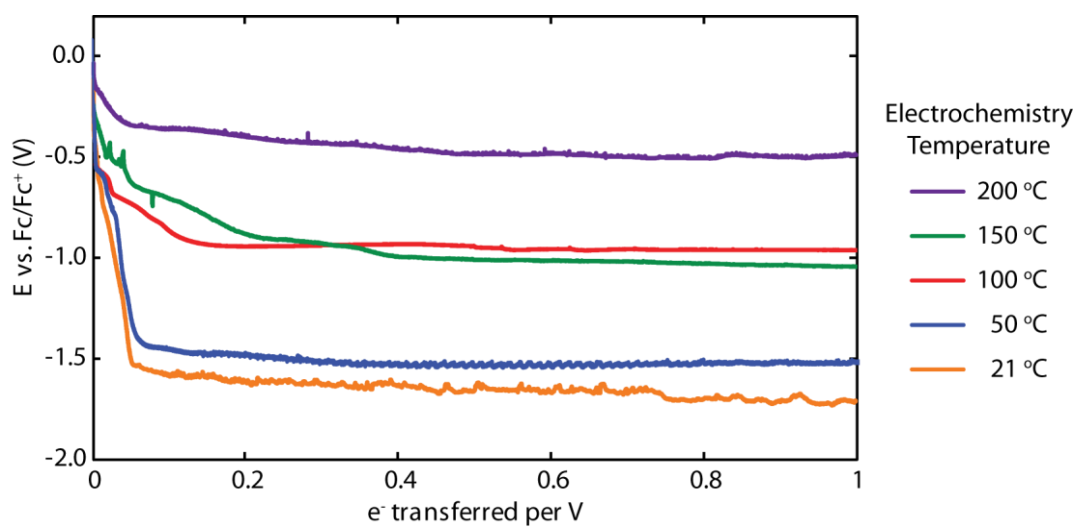

*Figure 30: Voltage profiles of electrochemically hydrogenated $VO_2$ at variable temperature.*

## 17 Electrochemical Metallisation at 200 °C

Electrochemical metallisation of $VO_2$ at 200 °C affords less hydrogenation than at 150 °C and moreover yields pure O1a, as shown by the XRD pattern (Figure 11) and $^1$H NMR spectrum (main text, Figure 5c). However, rather than the bulk susceptibility being dominated by the Pauli paramagnetism of the metallic O1a phase, with little of the Curie paramagnetism associated with the O1b phase, the susceptibility actually shows a decrease in the Pauli paramagnetism and the largest Curie constant of all the samples (Figure 26 and main text Figure 5e). This suggests that the O1a phase now contains localised paramagnetic defects.

The $^{17}$O NMR spectra recorded after electrochemical metallisation of an isotopically enriched sample at 200 °C support this conclusion (Figure 27). A signal is observed at ∼−800 ppm which exhibits a Curie–Weiss temperature dependence (Figure 29b), further suggesting the presence of paramagnetic defects in the O1a phase; the paramagnetic shift is negative due to the polarisation mechanism and becomes less negative with increasing temperature.

For this sample, the paramagnetic defects can only be observed in the $^{17}$O NMR spectrum, because paramagnetic vanadium centres relax too quickly to be observed via $^{51}$V NMR, and the hydrogen must be in a metallic environment, rather than being in the vicinity of a defect, because the $^1$H NMR spectrum exhibits a Knight shift. Nevertheless, the paramagnetic centres dominate the bulk magnetic susceptibility measurements, rather than the metallic contribution; the greater density of paramagnetic defects in this sample also explains the higher Néel temperature observed in the bulk magnetic susceptibility (Figure 26).

## 18 $^1$H NMR of Pristine $VO_2$ Thin Film

To determine whether the signal observed in the electrolyte gated $VO_2$ thin film was due to the gating process, the $^1$H NMR was recorded for a 200 nm $VO_2$ thin film without electrolyte gating, again lightly hand grinding the film to pack it into the sample rotor. Figure 31 shows the conventional and $T_1$-filtered $^1$H NMR spectra: unfortunately, there is a greater diamagnetic contamination of the pristine film which cannot be fully suppressed in the $T_1$-filtered spectrum and prevents direct comparison with the gated sample; however, no signal can be observed at 115 ppm in the $T_1$ filtered spectrum of the pristine $VO_2$, supporting the assertion that this signal in the gated $VO_2$ film is due to intercalated hydrogen in a metallic environment.

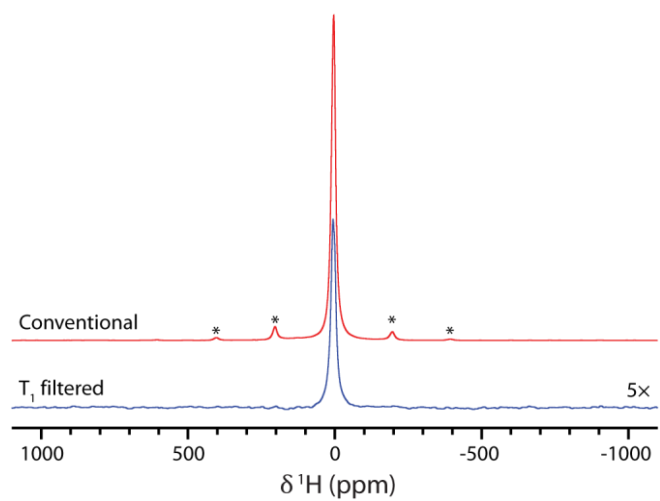

Figure 31: $^1$H NMR spectra of a 200 nm $VO_2$ thin film on 0.5 mm $TiO_2$ without electrolyte gating, recorded at 4.70 T and 40 kHz MAS, with spinning sidebands marked with asterisks. The conventional spectrum was obtained with a recycle delay of 0.05 s using a DEPTH pulse sequence[20] and subtracting the background. The $T_1$-filtered spectrum was recorded by taking the *difference between background subtracted spectra with recycle delays of 0.05 s and 0.1 s, scaling the spectra so as to remove as much as possible the diamagnetic signals.*

## 19 Deuteration of EMIm TFSI

Figure 32 shows the $^1$H NMR spectrum of pristine EMIm TFSI and the $^1$H and $^2$H NMR spectra after stirring in D$_2$O, showing selective deuteration of the carbene hydrogen at 8.3 ppm.[21] Comparison of the integrations of the $^1$H signals gives a deuteration of the carbene hydrogen of around 90 at%.

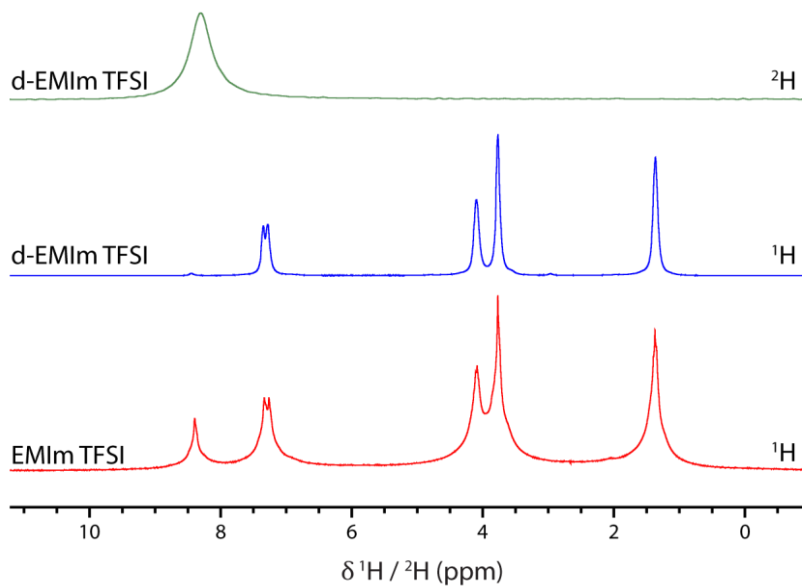

*Figure 32: The $^1$H NMR spectrum of pristine EMIm TFSI and the $^1$H and $^2$H NMR spectra after stirring in D$_2$O and subsequent drying, showing selective deuteration of the carbene hydrogen at 8.3 ppm. Spectra were recorded static at 7.05 T with a single-pulse experiment.*

## 20 $^2$H NMR of D$_x$VO$_2$ at Different Fields

Figure 33 shows the $^2$H NMR spectra of VO$_2$ electrochemically hydrogenated at 100 °C with deuterated EMIm TFSI, recorded at different magnetic field strengths. At lower fields the sideband manifold is dominated by the axial quadrupolar tensor, indicative of an axial bonding environment, and at higher fields the sideband manifold is dominated by a shift anisotropy.

The spectra were also modelled at each field by combining the quadrupolar tensor with a chemical shift anisotropy (CSA) tensor, the parameters for which are shown in Table 5. These parameters were found by fitting the quadrupolar tensor at 4.70 T, the CSA tensor at 11.75 T and then the Euler angles ($\alpha$, $\beta$, $\gamma$) relating these two tensors at 7.05 T, before iteratively minimising the difference at each field to yield a consistent fit. The solution is not perfect and may not be unique as changing the relative orientations of the tensors can lead to large and nonmonotonic changes in the sideband manifold, which are difficult to search using a local-optimisation routine. Nevertheless, the fit is reasonably consistent between fields and reproduces the major features, so the tensor parameters are likely to at least have the right order of magnitude.

The quadrupolar frequencies ($\nu_Q$) of O1a and O1b were found to be ~270 kHz and ~230 kHz respectively; this suggests that the hydrogen bonding of the H/D atoms to other oxygen atoms in VO$_2$ is reasonably weak, as hydrogen bonding reduces the electric field gradient,[22,23] with slightly stronger hydrogen bonding in the O1b phase. Furthermore, as stronger hydrogen bonding would require a greater distortion to bring the oxygen atoms closer together, this means that a large distortion around the intercalated hydrogen atoms would not be expected in H$_x$VO$_2$.

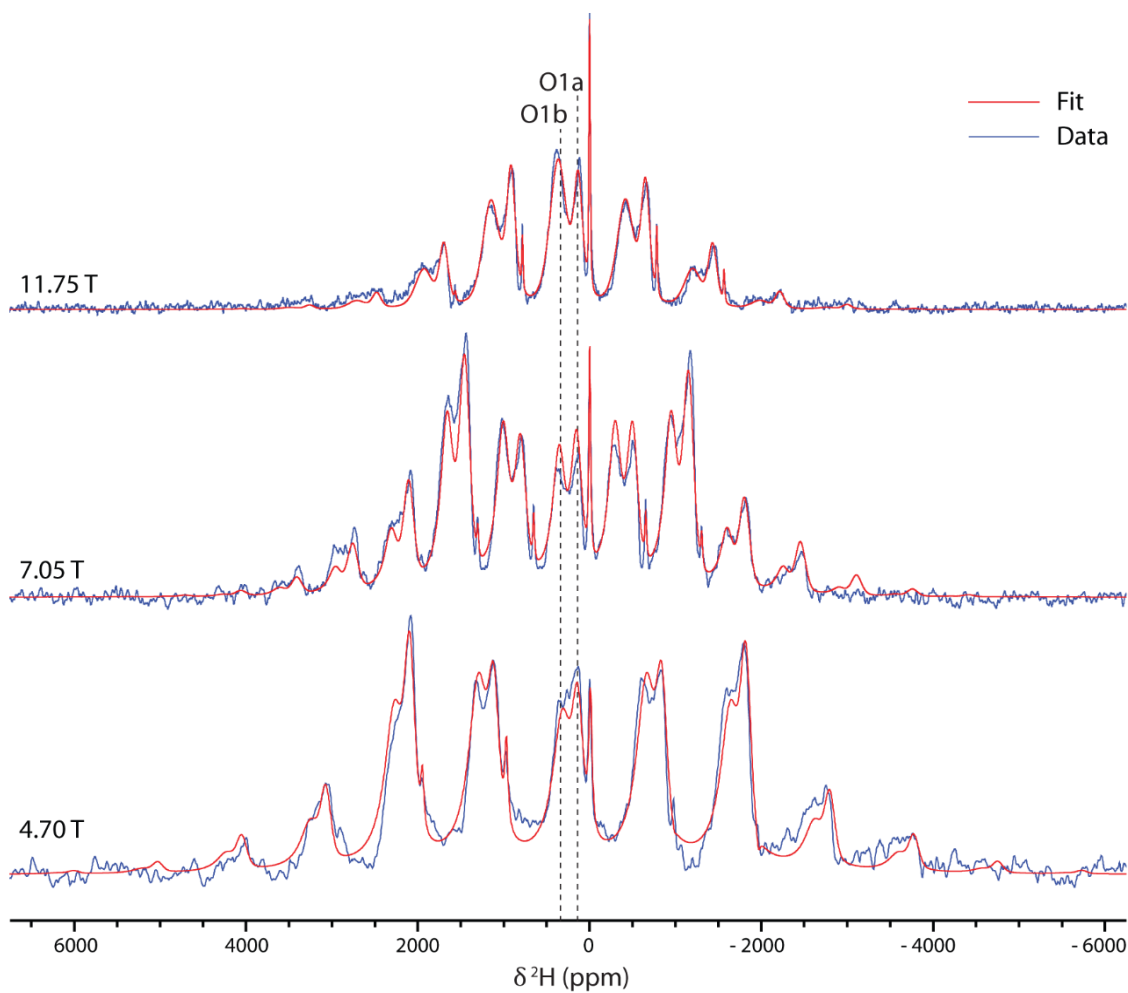

Figure 33: $^2$H NMR spectra of $VO_2$ after electrochemical metallisation at 100 °C with deuterated EMIm TFSI, recorded at fields of 4.70 T, 7.05 T and 11.75 T using a Hahn echo pulse sequence. The spectra at 4.70 T and 7.05 T were recorded at 30 kHz MAS, the spectrum at 11.75 T was recorded at 60 kHz MAS.

Table 5: Fitted chemical shift anisotropy (CSA) and asymmetry (η), quadrupolar frequency ($v_Q$) and asymmetry ($\eta_Q$), and Euler angles relating the two tensors (α, β, γ), for the deuterium environments in $D_xVO_2$.

|  | CSA /ppm | η | $v_Q$/kHz | $\eta_Q$ | α /° | β /° | γ /° |
|---|---|---|---|---|---|---|---|
| O1a | −1750 | 0.86 | 270 | 0.00 | 0 | 45 | 0 |
| O1b | 1410 | 0.99 | 230 | 0.00 | 0 | 45 | 0 |

## 21 Electrochemical Reduction with EIm TFSI and EM$_2$Im TFSI

Electrochemical metallisation experiments on $VO_2$ were also performed using different imidazolium-based ionic liquids: the more protic 1-ethylimidazolium bis(trifluoromethylsulfonyl)imide (EIm TFSI, Io-li-tec, 98%) and the less protic 1-ethyl-2,3-dimethylimidazolium bis(trifluoromethanesulfonyl)imide (EM$_2$Im TFSI, Tokyo Chemical Industry UK Ltd., 98%) (Figure 34). Similar degrees of hydrogenation were observed for all three ionic liquids (Table 6), however there is a difference in the voltage profiles observed during the electrochemical reduction (Figure 35): the acidity of the cations increases in the order EM$_2$Im < EMIm < EIm, and a less negative potential is observed following the same trend. At first glance the EM$_2$Im cation does not appear to have available protons, however it is possible to deprotonate the methyl group to form an N-heterocylic olefin.[24] The differences observed in the degree of hydrogenation are ascribed to small differences between the energy of the hydrogenation reaction and the limiting side reaction(s).

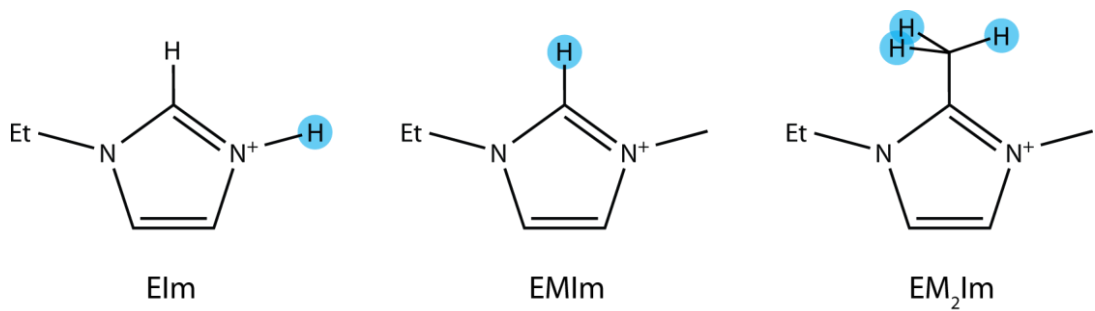

Figure 34: Different imidazolium-based ionic liquid cations, with the most acidic proton highlighted in each case.

Table 6: Degree of hydrogenation ($x$ in $H_xVO_2$, quantitative $^1H$ NMR) and phases present (XRD) for $VO_2$ electrochemically metallised at room temperature with different ionic liquids.

|  | EIm | EMIm | EM$_2$Im |
| --- | --- | --- | --- |
| $x$ | 0.049(3) | 0.037(2) | 0.091(5) |
| Phases | M + O1a | M | M + O1a |

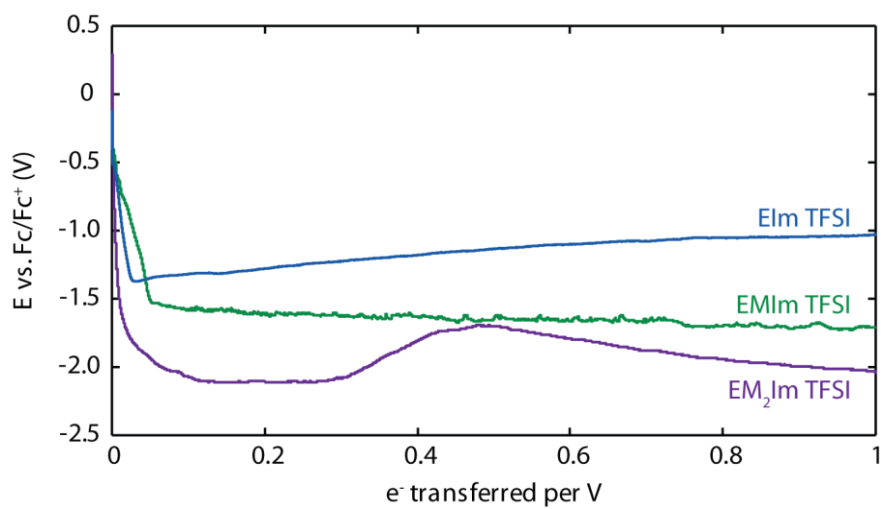

Figure 35: Voltage profiles for $VO_2$ electrochemically reduced at room temperature with different imidazolium-based ionic liquids.

## 22 Electrochemical Reduction with DEME TFSI

Another ionic liquid previously used in electrolyte gating experiments is diethylmethyl(2-methoxyethyl)ammonium bis(trifluoromethylsulfonyl)imide (DEME TFSI, cation shown in Figure 36a). Electrochemical reduction of bulk $VO_2$ with DEME TFSI also results in hydrogenation, as shown by $^1H$ NMR (Figure 36c). The DEME cation cannot form a stabilised N-heterocyclic carbene or olefin, as for imidazolium-based ionic liquids, so deprotonation must proceed by a different mechanism: one possibility is elimination of either the ethyl or methoxyethyl groups to form a neutral tertiary amine and ethene or methyl vinyl ether, respectively (Figure 36b, blue and green mechanisms); testing these mechanisms will be the subject of future work.

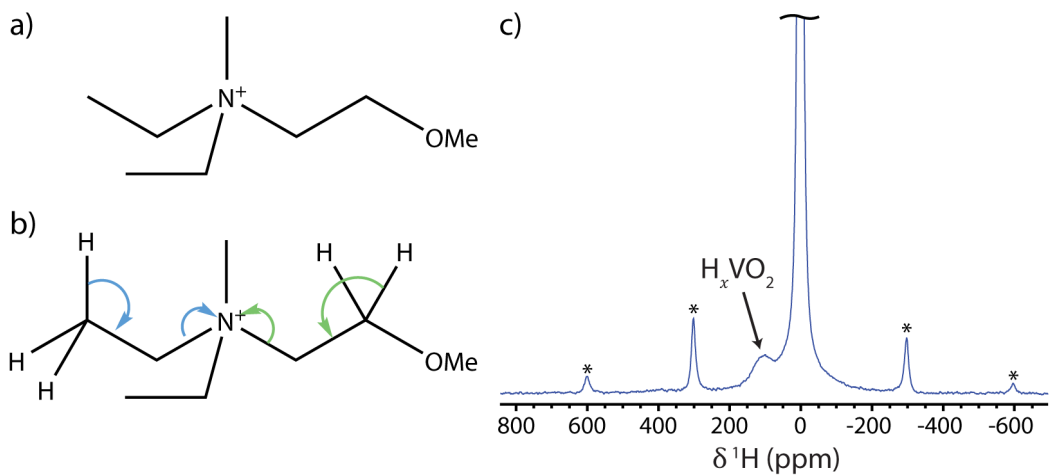

*Figure 36: a) The DEME cation. b) Two possible elimination mechanisms for deprotonation (green and blue arrows respectively). c) $^1$H NMR spectrum of $VO_2$ electrochemically reduced in DEME TFSI, recorded at 60 kHz MAS and 4.70 T with a Hahn echo pulse sequence. Sidebands are marked with asterisks.*

## 23 Electrochemical Reduction as a Function of Potential

To determine the onset voltage of $VO_2$ hydrogenation, bulk $VO_2$ composite films were galvanostatically reduced using EMIm TFSI until the potential reached a specified voltage relative to the reference electrode, before being held at this voltage for 24 hours; Figure 37 shows the degree of hydrogenation determined by *ex-situ* quantitative $^1$H NMR, as a function of this limiting voltage. There is not a sharp step in hydrogenation, but rather hydrogenation increases progressively with more negative voltage, which is consistent with a solid-solution rather than a two-phase reaction, before plateauing at ca. −0.75 V when some side-reaction prevents further hydrogenation. This is a less negative potential than in the galvanostatic experiment (ca. −1.6 V, see main text) because the overpotential is decreased when the potential is fixed and current allowed to respond, compared to when a fixed (higher) current is applied. The maximum amount of hydrogenation observed, $x$=0.021, is slightly lower than for galvanostatic reduction ($x$=0.037, see main text); this is ascribed to differences between batches of $VO_2$ composite films, most likely in particle size.

The onset voltage of $VO_2$ hydrogenation, ca. −0.5 V vs. Fc/Fc$^+$, is significantly less negative than the reported cathodic stability of EMIm TFSI (ca. −2.5 V vs. Fc/Fc$^+$);[25] however, this is not surprising since electrolyte stability is highly dependent on the electrode against which it is measured. Although EMIm TFSI is stable down to a low voltage against an inert glassy carbon electrode, $VO_2$ catalyses the breakdown which can therefore happen at a much less negative voltage. This can clearly be seen in Figure 21a, when the current is reversed the cathodic limit on the Pt electrode is more than a volt more negative than it was on the $VO_2$ electrode. Furthermore, because the $VO_2$ partakes in the reaction, the change in free energy is not simply that required to decompose the electrolyte but is lowered by the free energy released by reducing the $VO_2$, and the voltage is thus correspondingly less negative.

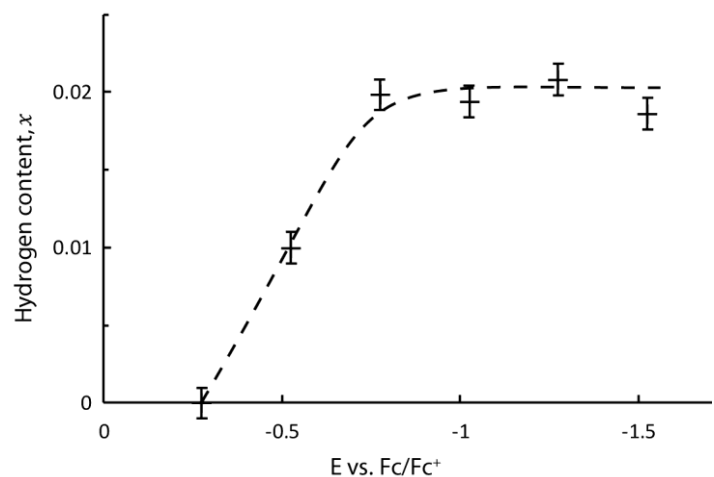

*Figure 37: Hydrogen content $x$, from quantitative $^1$H NMR, as a function of limiting voltage for electrochemical reduction of bulk VO$_2$ composite films with EMIm TFSI.*

# 24 Variable Temperature XRD of M + O1a

To determine the effect of heating to above the MIT temperature on the two-phase M + O1a region of the phase diagram, variable temperature XRD patterns were recorded for VO$_2$ electrochemically metallised at 50 °C, $x \approx 0.1$. XRD patterns were recorded at room temperature, 100 °C, and then room temperature again, using a heated sample holder built in-house, in reflection mode, with a beryllium window. Figure 38 shows two expanded sections of the XRD patterns: in both room temperature patterns, there are two reflections around $2\theta = 28°$ corresponding to the M and O1a phases, which coalesce into the single rutile reflection at 100 °C. The same effect is observed in the $2\theta = 54° - 58°$ region, and in particular the orthorhombic peak at $2\theta = 54.7°$ is not present in the 100 °C pattern. This shows that the two-phase M + O1a region of the phase diagram at room temperature becomes a single rutile (R) phase above the MIT, due to a greater solubility of hydrogen in the rutile phase compared to the monoclinic phase. The M and O1a phases are again observed after cooling down to room temperature, indicating that this phase transition is reversible.

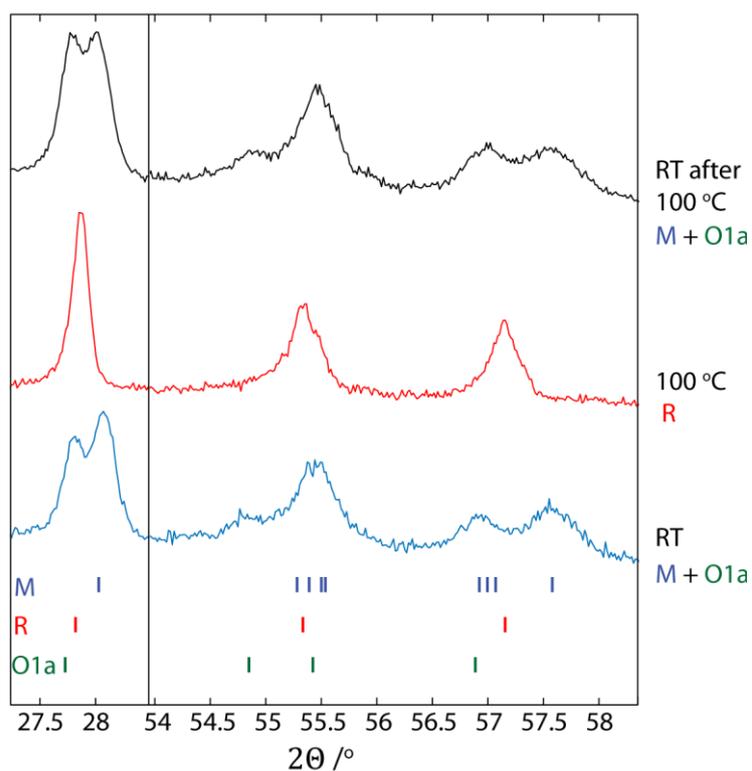

*Figure 38: Selected regions of the XRD patterns of VO$_2$ electrochemically metallised at 50 °C, recorded at room temperature, 100 °C, and then room temperature again.*

# 25 References

(1) De Souza, A. C.; Pires, A. T. N.; Soldi, V. Thermal Stability of Ferrocene Derivatives and Ferrocene-Containing Polyamides. *J. Therm. Anal. Calorim.* **2002**, *70* (2), 405–414.

(2) Liebsch, A.; Ishida, H.; Bihlmayer, G. Coulomb Correlations and Orbital Polarization in the Metal-Insulator Transition of VO2. *Phys. Rev. B* **2005**, *71* (8), 085109.

(3) Eyert, V. VO2: A Novel View from Band Theory. *Phys. Rev. Lett.* **2011**, *107* (1), 2–5.

(4) Brito, W. H.; Aguiar, M. C. O.; Haule, K.; Kotliar, G. Metal-Insulator Transition in VO2: A DFT + DMFT Perspective. *Phys. Rev. Lett.* **2016**, *117* (5), 056402.

(5) Clark, S. J.; Segall, M. D.; Pickard, C. J.; Hasnip, P. J.; Probert, M. J.; Refson, K.; Payne, M. C. First Principles Methods Using {CASTEP}. *Z. Krist.* **2005**, *220*, 567–570.

(6) Pickard, C. J.; Mauri, F. All-Electron Magnetic Response with Pseudopotentials: {NMR} Chemical Shifts. *Phys. Rev. B* **2001**, *63*, 245101.

(7) Yates, J. R.; Pickard, C. J.; Mauri, F. Calculation of NMR Chemical Shifts for Extended Systems Using Ultrasoft Pseudopotentials. *Phys. Rev. B* **2007**, *76*, 24401.

(8) Profeta, M.; Mauri, F.; Pickard, C. J. Accurate First Principles Prediction of 17 O NMR Parameters in SiO 2 : Assignment of the Zeolite Ferrierite Spectrum. *J. Am. Chem. Soc.* **2003**, *125* (2), 541–548.

(9) Perdew, J. P.; Burke, K.; Ernzerhof, M. Generalized Gradient Approximation Made Simple. *Phys. Rev. Lett.* **1996**, *77* (18), 3865–3868.

(10) MacKenzie, K. J.; Smith, M. E. *Multinuclear Solid-State Nuclear Magnetic Resonance of Inorganic Materials*; 2002.

(11) Geller, S.; Romo, P.; Remeika, J. P. Refinement of the Structure of Scandium Sesquioxide. *Zeitschrift für Krist. Mater.* **1967**, *124* (1–6), 136–142.

(12) Restori, R.; Schwarzenbach, D.; Schneider, J. R. Charge Density in Rutile, TiO2. *Acta Crystallogr. Sect. B Struct. Sci.* **1987**, *43* (3), 251–257.

(13) Bachmann, H.G.; Ahmed, F.R.; Barnes, W. H. The Crystal Structure of Vanadium Pentoxide. *Zeitschrift fuer Krist. Krist. Krist. Krist.* **1961**, *115*, 110–131.

(14) Longo, J. M.; P, K. A Refinement of the Structure of VO2. *Acta Chem. Scand.* **1970**, *24* (2), 420.

(15) Bonhomme, C.; Gervais, C.; Babonneau Florenceand Coelho, C.; Pourpoint, F.; Azais, T.; Ashbrook, S. E.; Griffin, J. M.; Yates, J. R.; Mauri, F.; Pickard, C. J. First-Principles Calculation of NMR Parameters Using the Gauge Including Projector Augmented Wave Method: A Chemist's Point of View. *Chem. Rev.* **2012**, *112*, 5733.

(16) Chippindale, A. M.; Dickens, P. G.; Powell, A. V. Synthesis, Characterization, and Inelastic Neutron Scattering Study of Hydrogen Insertion Compounds of VO2(Rutile). *J. Solid State Chem.* **1991**, *93* (2), 526–533.

(17) Thurber, K. R.; Tycko, R. Measurement of Sample Temperatures under Magic-Angle Spinning from the Chemical Shift and Spin-Lattice Relaxation Rate of 79Br in KBr Powder. *J. Magn. Reson.* **2009**, *196* (1), 84–87.

(18) Mydosh, J. A. *Spin Glasses: An Experimental Introduction*; Taylor & Francis: London, 1993.

(19) Carlier, D.; Ménétrier, M.; Grey, C. P.; Delmas, C.; Ceder, G. Understanding the NMR Shifts in Paramagnetic Transition Metal Oxides Using Density Functional Theory Calculations. *Phys. Rev. B* **2003**, *67* (17), 174103.

(20) Robin Bendall, M.; Gordon, R. E. Depth and Refocusing Pulses Designed for Multipulse NMR with Surface Coils. *J. Magn. Reson.* **1983**, *53* (3), 365–385.

(21) Fujii, K.; Soejima, Y.; Kyoshoin, Y.; Fukuda, S.; Kanzaki, R.; Umebayashi, Y.; Yamaguchi, T.; Ishiguro, S.; Takamuku, T. Liquid Structure of Room-Temperature Ionic Liquid, 1-Ethyl-3-Methylimidazolium Bis-(Trifluoromethanesulfonyl) Imide. *J. Phys. Chem. B* **2008**, *112* (14), 4329–4336.

(22) Kim, G.; Blanc, F.; Hu, Y. Y.; Grey, C. P. Understanding the Conduction Mechanism of the Protonic Conductor CsH2PO4 by Solid-State NMR Spectroscopy. *J. Phys. Chem. C* **2013**, *117* (13), 6504–6515.

(23) Blinc, R.; Hadži, D. Deuteron Quadrupole Coupling and Hydrogen Bonding in Crystals. *Nature* **1966**, *212* (5068), 1307–1309.

(24) Saptal, V. B.; Bhanage, B. M. N-Heterocyclic Olefins as Robust Organocatalyst for the Chemical Conversion of Carbon Dioxide to Value-Added Chemicals. *ChemSusChem* **2016**, *9* (15), 1980–1985.

(25) Mousavi, M. P. S.; Dittmer, A. J.; Wilson, B. E.; Hu, J.; Stein, A.; Bühlmann, P. Unbiased Quantification of the Electrochemical Stability Limits of Electrolytes and Ionic Liquids. *J. Electrochem. Soc.* **2015**, *162* (12), A2250–A2258.